\documentclass[onecolumn,numberedappendix,appendixfloats,iop,apj]{emulateapj}
\usepackage{graphicx}
\usepackage{epstopdf}
\usepackage{mathrsfs}
\usepackage{amssymb}
\usepackage[title]{appendix}
\usepackage{amsmath}
\usepackage{color}
\usepackage{bm}

\newcommand{\pri}{\prime}
\newcommand{\pr}{\prime}

\newcommand{\vu}{\mathbf u}

\newcommand{\vrr}{\mathbf r}

\newcommand{\vx}{\mathbf x}
\newcommand{\vk}{\mathbf k}
\newcommand{\vq}{\mathbf q}
\newcommand{\vp}{\mathbf p}

\newcommand{\vA}{\bm A}
\newcommand{\vmA}{\bm {\mathcal A}}

\newcommand{\mH}{\mathcal H}
\newcommand{\mA}{\mathcal A}
\newcommand{\mP}{\mathcal P}
\newcommand{\mZ}{\mathcal Z}
\newcommand{\mK}{\mathcal K}

\newcommand{\drho}{\delta_{\rho}}
\newcommand{\Drho}{\Delta_{\rho}}
\newcommand{\mD}{\mathcal{D}}

\newcommand{\vPsi}{\bm \Psi}
\newcommand{\vGamma}{\bm \Gamma}
\newcommand{\vpsi}{\bm \psi}
\newcommand{\vchi}{\bm \chi}
\newcommand{\vChi}{\bm X}

\newcommand{\vX}{\bm X}
\newcommand{\vY}{\bm Y}
\newcommand{\vy}{\bm y}

\newcommand{\Chi}{X}

\newcommand{\vE}{\bm E}
\newcommand{\vOmega}{\bm \Omega}

\newcommand{\vlambda}{\bm \lambda}

\newcommand{\hr}{\hat{r}}
\newcommand{\hk}{\hat k}

\def\presuper#1#2%
  {\mathop{}%
   \mathopen{\vphantom{#2}}^{#1}%
   \kern-\scriptspace%
   #2}

\begin{document}

\title{Statistical Decoupling of Lagrangian Fluid Parcel in Newtonian Cosmology}

\author{Xin Wang\altaffilmark{1,2,3}, Alex Szalay\altaffilmark{1}}
\altaffiltext{1}{Department of Physics \& Astronomy, Johns Hopkins University, 
Baltimore, MD, 21218, USA}
\altaffiltext{2}{Canadian Institute for Theoretical Astrophysics, 60 St. George St., Toronto, ON,
M5H 3H8, Canada}
\altaffiltext{3}{xwang@cita.utoronto.ca}

\label{firstpage}

\begin{abstract}
The Lagrangian dynamics of a single fluid element within a self-gravitational  matter field is 
intrinsically non-local due to the presence of the tidal force. 
This complicates the theoretical investigation of the non-linear evolution of various cosmic objects, 
e.g. dark matter halos, in the context of Lagrangian fluid dynamics, since a fluid parcel with given 
initial density and shape may evolve differently depending on their environments. 
In this paper, we provide a statistical solution that could decouple this environmental dependence. 
After deriving the probability distribution evolution equation of the matter field, our method 
produces a set of closed ordinary differential equations whose solution is uniquely determined by 
the initial condition of the fluid element. 
Mathematically, it corresponds to the projected characteristic curve of the transport equation 
of the density-weighted probability density function ($\rho$PDF). 
Consequently it is guaranteed that the one-point $\rho$PDF would be preserved by evolving 
these local, yet non-linear, curves with the same set of initial data as the real system.
Physically, these trajectories describe the mean evolution averaged over all environments by
substituting the tidal tensor with its conditional average. 
For Gaussian distributed dynamical variables, this mean tidal tensor 
is simply proportional to the velocity shear tensor, and the dynamical system would recover the 
prediction of Zel'dovich approximation (ZA) with the further assumption of the linearized 
continuity equation. 
For Weakly non-Gaussian field, the averaged tidal tensor could be expanded perturbatively
as a function of all relevant dynamical variables whose coefficients are determined by 
the statistics of the field. 
\end{abstract}

\keywords{cosmology: theory, dark matter, large-scale structure of universe}

\maketitle

\section{Introduction}

The large-scale structure of the Universe encodes valuable information of various physical 
processes including primordial physics, non-linear gravitational dynamics, and late-time baryonic
galaxy formation. 
While the primordial and baryonic physics would require more complicated theoretical treatment,
the gravitational effect could be studied in the much simpler context of Newton's theory 
of gravity, together with either the equations of motion for individual dark matter particles 
\citep{P80,Bert95} or the fluid conservation law \citep{P80,BCGS02}.
This has lead to tremendous progresses in structure formation theory in the last several decades. 
Especially, non-linear perturbation theory (PT), both Eulerian and Lagrangian \citep{BCGS02}, 
have evolved into sophisticated forms \citep{CS06a,CS06b,BCS08,M08a,TH08,P08} 
that could provide a very accurate estimation of the matter clustering in the weakly non-linear 
regime. 
In the deeply non-linear region, however, PT performs poorly. Alternatively, one could try to 
decouple the evolution of highly non-linear objects, e.g. dark matter halos, from their surroundings 
and study separately. 
This approach leads to phenomenological theory like the halo model \citep{NS52,CS02}. 
In such model, these viralized halos provide the environment for most of the observed 
galaxies \citep{WR78,CS02}; connect the statistical properties of galaxies with their spatial 
and size distribution.  
Consequently, it has become very popular for its intuitive simplicity and wide range of applications.

However, the analytical investigation of these highly non-linear objects has been complicated by
one intrinsic property of the Newtonian gravity -- the non-locality, which means that
the gravitational effect felt by a single element is determined by the matter distributed in the entire 
Universe, and that the whole field has be to taken into account for realistic modeling. 
Therefore, two proto-halos with exact the same initial density and shape might evolve into
very different states cased by their distinct environments.  
Furthermore, for various purposes, we are also interested in other cosmic-web 
morphologies \citep{BKP96,LK99} including filaments, walls and voids
\citep{ST04,ACG05,WZ05,BE07,MD07,WCW07,HP07a,HC07b}. 
The same problem arises when we concentrate on discrete matter patches and
attempt to understand their Lagrangian evolutions\footnote{The Lagrangian 
evolution mentioned here concerns individual fluid patches, so is different from the
Lagrangian perturbation theory which still deals with the field. } as well.   
From the dynamical point of view, this means the evolution equation is not closed
except the perfect symmetric system like the spherical collapse (SC) model
\citep{GG72,PS74} in which the tidal effect vanishes, 
and we are trying to describe the system with only a limited number of dynamical degree
of freedoms.

Then how to build a theoretical framework in the Newtonian cosmology
for those non-linear objects with insufficient number of degree of freedoms? 
First of all, as argued by \cite{BM96}, the gross features of a halo are sensitive to only a few
dynamical variables of the progenitor, such as the bulk velocity, tidal tensor and average over density. 
The model then proposed is the homogenous ellipsoidal collapse (HEC) model, which provides 
much more information than SC, and has been incorporated in many frameworks like the 
peak-patch approach \citep{BM96} and the excursion set model \citep{ST02}. 
Therefore, to a certain extent, we could approximate internal dynamics with just the density and 
shape. 
For the external tidal effect, however, HEC then assumes the linear approximation, believing that the 
internal dynamics is well decoupled. Clearly, this would not accurately address the 
environmental dependence from the surrounding, and it is this external contribution we are going to 
investigate in this paper.

The Lagrangian fluid dynamics of single fluid element basically has the same number of internal 
degree of freedoms as the homogenous ellipsoidal collapse model. 
As a result, it has long been proposed as alternative theoretical framework for investigating the 
highly non-linear process of halo collapse \citep{BJ94}.
Most importantly, in the general relativity, the Lagrangian dynamical system is actually closed. 
Pioneered by  \cite{Eh61} and \cite{KT61}, then applied in cosmological perturbation theory by 
\cite{H66} and \cite{E71},  the Lagrangian fluid dynamics is fully described by a set of coupled differential 
equations for fluid density, velocity divergence, shear, vorticity and the Weyl tensor, which could 
further be decomposed into electric and magnetic parts.  
However, the Newtonian version of the theory did not arrive trivially, as the evolution 
equation of the tidal tensor, i.e. the Newtonian counterpart of electric Weyl tensor, is 
missing and the magnetic part is not even defined \citep{E71}. 
Motivated by the latter statement, \cite{MPS93} showed that a vanishing magnetic Weyl 
tensor would lead to a set of local Lagrangian fluid equations 
\citep{MPS94,BMP95,LDE95}, known as the `silent universe' model since the 
tensorial interaction between neighboring elements is neglected. 
\cite{BJ94} then demonstrated that, in this model, the filamentary collapse is favored instead of
the pancakes, as predict by ZA. 
Soon, \cite{KP95} criticized this model as an incorrect Newtonian limit of the relativistic 
equations and demonstrated the proper tidal evolution equation should contain 
non-local source terms as well. 
Meanwhile, the generalization to non-vanishing magnetic Weyl tensor has also 
triggered some further discussions \citep{BH94,ED97}, 
but `it remains an unsolved problem to show satisfactorily how the non-linear 
Newtonian versions of the equations can be derived in a suitable limit from the 
relativistic theory' \citep{EMM12}.

Although unsuccessfully, these efforts originate from the more fundamental general relativity, 
and clearly suggest the importance of appropriate incorporating the non-local tidal effect. 
Besides, the resulting dynamical system are fully non-linear ordinary differential equations 
(ODEs), and could be easily solved. 
One natural approach is to supplement the full evolution equation of the tidal tensor 
or the magnetic Weyl tensor. 
In this spirit, \cite{HB96} tried to close the system by imposing local approximation for 
the evolution of tidal tensor. 
However, the meaning of these local approximations is not clear.
Given initial density and shape of an element, the non-local tidal tensor would lead to various
trajectories depending on the specific location in the Universe. 
Therefore, the underlying question behind the local approximation is how to select one out of 
those trajectories. 
Practically, given some experimental or simulated data, and assuming that one could follow fluid 
elements all the way down to desired redshift, the most obvious solution is to simple take some mean 
among all trajectories with given initial condition. 
With that said, a conceptually well-defined procedure of local approximation will 
inevitably be statistical.

Hence in this paper, to address these issues, we are going to present a statistical method that 
is well developed in turbulence, known as the probability density function (PDF) based method
\citep{Pop85,WM14},  where the key ingredient is the evolution of one-point PDF. 
In the cosmological context, the one-point PDF of gravitational evolved field is, 
in particular, interesting.
Based on the simulation measurement, the multi-point density PDF could be well 
approximated by a Gaussian N-point part, namely the copula,  together with a 
non-Gaussian one-point PDF\citep{SBMM10}. 
It suggests that a large amount of gravitational non-linear information is stored in the 
one-point  statistics. 
Consequently, a logarithmic transform \citep{NSS09} or a local Gaussianization 
\citep{NSS11} of the density field would then be able to retrieve abundant information 
from the standard two-point correlation function, which otherwise would  leak into higher-order 
statistics.

This paper is organized as follows. 
After a brief review of the Lagrangian dynamics in section 2, we introduce our main principle and 
formalism of the effective Lagrangian evolution in section 3.  
In section 4, we first perform a Gaussian closure for tidal tensor, and then extend the
calculation into weakly non-Gaussian region. 
Finally in section 5, we discuss the generalization of the method to incorporate stochasticity, 
and then the necessity of going beyond the mean evolution. 
To avoid distraction, we leave the details of the evolution equation of density-weighted
PDF in Appendix A, and of the formula related to the statistical closure in Appendix B and C. 
In Appendix D, we perform an alternative calculation for Gaussian closure.

\section{The Lagrangian Fluid Dynamics and Non-locality}

In the Newtonian cosmology, the gravitational instability of the large-scale
structure is described by the fluid conservation equation of density contrast $\drho$
and peculiar velocity $\vu$.  Before shell-crossing, they satisfy the continuity and 
Eulerian equations respectively. 
In Lagrangian fluid dynamics, one denotes the total derivative as 
$d/d\tau = \partial/\partial \tau + \vu\cdot \nabla $,  so these two equations are 
\begin{eqnarray}
\label{eqn:Lagrangian_dyn_details}
\frac{d}{d\tau} \drho + (1+\drho) \theta  = 0,  \qquad\qquad
\frac{d}{d\tau} u_i + \mH(\tau) u_i = - \Phi_i  , 
\end{eqnarray}
where $\tau$ is the conformal time, $\Phi_i = \nabla_i \Phi $ is the gradient of 
peculiar gravitational potential $\Phi$, which obeys the Poisson equation
\begin{eqnarray}
\label{eqn:Poisson_eq}
	\nabla^2 \Phi = 4\pi G_N \bar{\rho} a^2 \drho. 
\end{eqnarray}
Here $G_N$ is the gravitational constant, $a$ is scale factor, $\bar{\rho}$ is the mean
background density. 
Due to the presence of the velocity divergence $\theta=\nabla\cdot \vu$ in Equation 
(\ref{eqn:Lagrangian_dyn_details}), one also needs the Lagrangian equation of the 
spatial gradient of peculiar velocity $A_{ij} = \nabla_i u_j$, which could be derived by taking 
the gradient of Eulerian equation
\begin{eqnarray}
\label{eqn:Lagrangian_dyn_details_Aij}
\frac{d}{d\tau} A_{ij} + \mH(\tau)A_{ij} + A_i^k A_{kj} &=& -\Phi_{ij} . 
\end{eqnarray}
Here tensor $\Phi_{ij} = \nabla_{ij}\Phi$ is defined as the Hessian matrix of potential $\Phi$. 
It is more convenient to decompose $A_{ij}$ as
\begin{eqnarray}
A_{ij} = \frac{\theta}{3}\delta^K_{ij} + \sigma_{ij} + \omega_{ij}, 
\end{eqnarray}
where the trace part $\theta$ is divergence, $\sigma_{ij}$ is the traceless symmetric shear tensor, 
and $\omega_{ij}$ is the anti-symmetric vorticity tensor. 
Notice that we have also introduced the Kronecker delta function $\delta^K_{ij}$. 
Similarly, the tensor $\Phi_{ij}$ could be decomposed as
\begin{eqnarray}
\Phi_{ij}  = \frac{\nabla^2 \Phi}{3} \delta^K_{ij} + \varepsilon_{ij} 
= \frac{4\pi G_N \bar{\rho} a^2 \drho  }{3} \delta^K_{ij} + \varepsilon_{ij}. 
\end{eqnarray}
The trace part $\nabla^2 \Phi$ is proportional to the density contrast via Poisson equation, and 
the symmetric tidal tensor $\varepsilon_{ij}$ is the only traceless part here. 
Therefore, in the standard cosmological dust model with zero primordial vorticity, $\omega_{ij}$ 
remains zero before the shell-crossing. 
In summary, the full Lagrangian dynamics of a single fluid element reads as
\begin{eqnarray}
\label{eqn:dyn_eqn_full}
\frac{d}{d\tau} \drho  &=& - (1+\drho) \theta ,  \nonumber \\
\frac{d}{d\tau} \theta &=& -\left [ \mH(\tau)\theta + \frac{1}{3}\theta^2 + 
\sigma^{ij} \sigma_{ij}  \right] -4\pi G_N \bar{\rho} a^2 \drho,  \nonumber \\
\frac{d}{d\tau} \sigma_{ij}  &=& -\left[ \mH(\tau) \sigma_{ij} + \frac{2}{3}\theta\sigma_{ij}
+ \sigma_{ik}\sigma^k_j   - \frac{1}{3} \sigma_{mn}\sigma^{mn}  \delta^K_{ij} \right]  - \varepsilon_{ij}, 
\end{eqnarray}
In the following,  we will group all dynamical variables as 
$\vpsi = \{\drho, A_{ij}  \} =  \{\drho, \theta, \sigma_{ij}  \} $, so that Equation (\ref{eqn:dyn_eqn_full}) 
could be simplified as  
\begin{eqnarray}
\label{eqn:dyn_sys_tot}
\frac{d}{d\tau} \vpsi (\tau) = \vchi [\vpsi, \varepsilon_{ij}], 
\end{eqnarray}
where $ \vchi [\vpsi, \varepsilon_{ij}] $ is the function depends on $\vpsi$ and $\varepsilon_{ij}$. 
And we will use Greek letters for the indices of grouped vector
$\psi_{\alpha}$ while Latin letters for the spatial coordinates.

\begin{figure*}
\begin{center}
\includegraphics[width=\textwidth]{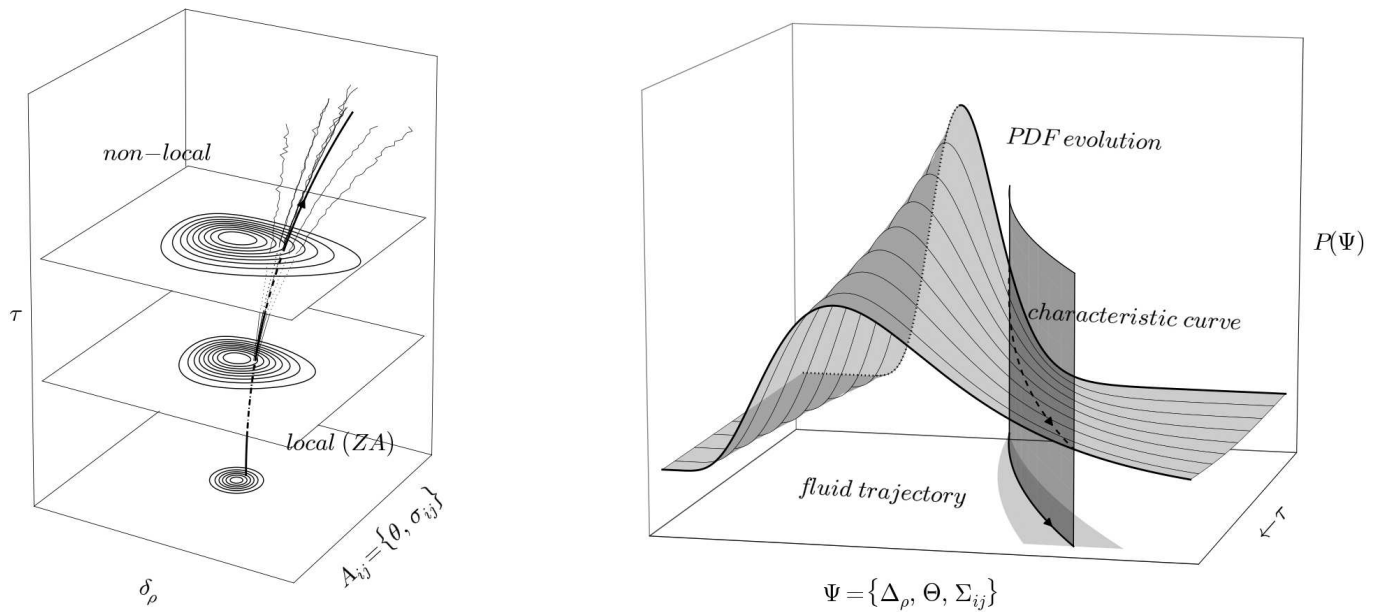}
\end{center}
\caption{\label{fig:PDF_illustration}
({\it Left}): The Lagrangian evolution of fluid elements in the parameter space of 
$\vpsi=\{\drho, \theta, \sigma_{ij} \}$. Contours corresponds to the evolution of one-point
probability density function, from Gaussian to some skewed distribution. 
At early stage, the evolution is fully described by Zel'dovich approximation, therefore is local, 
i.e. uniquely determined by initial condition $\vpsi(\tau_i)$. 
However, after entering into non-linear regime, the tidal tensor will not be a simple deterministic
function of $\vpsi$ locally, and therefore will deviate from and spread around local solutions. 
Any local approximation that attempts to close the dynamical system then corresponds to
either "selecting" one particular trajectory or taking some non-local operation over all relevant 
curves. In this paper, we are interested in finding some statistical mean trajectory given 
initial density and shape. 
({\it Right}):   Schematics of the PDF based method for deriving the mean trajectory in 
the parameter space of dynamical variables $\vPsi=\{ \Drho, \Theta, \Sigma_{ij} \}$ 
(notice that we have explicitly distinguished the dynamical variable $\vpsi$ of the real system 
from the sample space variable $\vPsi$). From the genuine 
dynamical system, we derive the evolution equation of one-point PDF (contours in the {\it left} panel), 
which as shown in the main text, is a first order partial differential equation. 
The characteristic curves, which serve
as standard method for solving this type of equation, is then projected back into the dynamical
space $\vPsi$. The projected curves that described by ordinary differential equation 
(\ref{eqn:effective_traj_Lag}) resembles the full dynamical system except replacing the 
non-local source term by its conditional average. 
With given initial condition $\vpsi(\tau_i)$, unlike genuine dynamical trajectories which resides somewhere 
within the grey band, the projected characteristic curve is uniquely determined by the initial condition.
}
\end{figure*}

It is clear that the only unclosed term is the tidal tensor $\varepsilon_{ij}$, which need to be solved from 
Poisson equation. 
In the early evolutionary stage, however, the large-scale structure was well described by the 
Zel'dovich approximation (ZA) \citep{ZA70}, where the particle displacement field is fully determined by the local
density.  The velocity potential is then proportional to the gravitational potential, 
so that the dynamical evolution of a fluid element is the same as equation (\ref{eqn:dyn_eqn_full})  
except replacing the tidal tensor by the shear tensor  \citep{HB96,BCGS02} 
\begin{eqnarray}
\varepsilon_{ij} = - \frac{4 \pi G_N a^2 \bar{\rho} }{\mH f} \sigma_{ij}. 
\end{eqnarray} 
In general, however, the peculiar tidal tensor at any Eulerian location $\vx$ is an integral over
the entire space 
\begin{eqnarray}
\varepsilon_{ij} (\vx) = G_N \bar{\rho} a^2 \int d^3x^{\pr} \left [ 
\frac{\delta^K_{ij}}{r^3} - 3 \frac{r_i r_j }{r^5} \right] \drho(\vx^{\pr}), 
\end{eqnarray}
which would also vary depending on the specific spatial location. 
Consequently, as shown in the left panel of Figure. (\ref{fig:PDF_illustration}), the trajectory of a fluid 
element in parameter space $\vpsi = \{ \drho, \theta, \sigma_{ij} \}$ will deviate from, and more 
importantly,  spread around the trajectory of ZA, and will not be uniquely determined by the initial 
condition $\vpsi(\tau_i)$. 
The goal of this paper, as already mentioned in the introduction, is then try to obtain a 
statistical meaningful trajectory that is uniquely determined by $\vpsi(\tau_i)$ (as presented
by the thick solid curve in the left panel of Figure (\ref{fig:PDF_illustration}) ).

\section{The Evolution of One-Point Statistics and the Mean Lagrangian Evolution}

In this section, we will demonstrate a statistical method to estimate the local mean trajectory (LMT). 
Specifically, by local, we mean that the trajectory is uniquely determined by the initial condition of the 
dynamical variable $\vpsi$, as in ZA. 
This will be achieved by discussing the one-point statistics evolution equation which is derived from 
basic dynamical equation (\ref{eqn:dyn_sys_tot}). 
Then, the mean trajectory will be obtained from the projection of the characteristic curves regarding
this partial differential equation of PDF.
Therefore, the methodology here exhibits a certain kind of spiral structure. 
From  the genuine dynamics to the PDF evolution, we lost the information of the individual trajectories. 
However, the remaining information leads to the mean evolution of all individual particles in a 
given realization.

\subsection{From Dynamics to Statistics}

Assume we could follow all fluid particles in a self-gravitational system, and we would like
to consider the probability of a random selected parcel (e.g. originally located at some Lagrangian 
position $\vq$) whose  dynamical variables $\vpsi$ at time $\tau$ equals $\vPsi$. 
Here, following the turbulence convention, 
the capitalized $\vPsi$ denotes the sample space variable corresponding to $\vpsi$
\begin{eqnarray}
\vPsi =\{ \Drho, \mA_{ij} \} = \{\Drho, \Theta, \Sigma_{ij} \}. 
\end{eqnarray}
Initially, the resulting probability density function is simply Gaussian, as will be assumed throughout the paper. 
As the dynamical system evolves following equation (\ref{eqn:dyn_sys_tot}), the PDF will then change gradually. 
In the following, we will term it the Lagrangian PDF ($\mP_L(\vPsi; \tau)$ \footnote{
In this paper, semicolon will be used to divide different types of arguments, whereas comma among the 
same type of arguments. For example, $\mP(x, y; \tau)$
means the joint probability density function of random variable $x$ and $y$, and it also depends on 
time $\tau$. }),  since the underlying dynamics is in the Lagrangian sense. 
For a single realization (or a randomly selected particle in the system), the probability density function, i.e. 
the fine-grained PDF ($\mP^{f}_L$),  could simply be written as Dirac-$\delta$ function \citep{Pop85}
\begin{eqnarray}
\label{eqn:PDF_fg_Lag}
\mP^{f}_L(\vPsi;\tau) = \delta_D\left [ \vpsi(\tau) - 
\vPsi\right]  = \delta_D\left [\drho(\tau) -\Drho \right]  
 \delta_D\left [ \vA(\tau) - \vmA \right], 
\end{eqnarray} 
since the probability of finding the dynamical variable $\vpsi=\vPsi$ at $\tau$ will only be non-zero at 
the value determined by the dynamical equation
 (\ref{eqn:dyn_sys_tot}). Here $\delta_D$ is the Dirac-$\delta$ function. 
Now let us take an ensemble of particles, e.g. all particles in the system. 
By definition, this fine-grained PDF then relates to  
$\mP_L(\vPsi;\tau)$ by taking the ensemble average  $\langle \cdot \rangle_L$ associated 
with $\mP_L(\vPsi;\tau)$
\begin{eqnarray}
\label{eqn:PL_rel}
\langle P^f_L (\vPsi; \tau) \rangle_L 
= \int d\vPsi^{\pri} P_L (\vPsi^{\pri}; \tau) \delta_D(\vPsi^{\pri}- \vPsi) = P_L (\vPsi; \tau). 
\end{eqnarray}
Practically, at any given time $\tau$, $\langle \cdot \rangle_L$ is achieved by averaging over all 
fluid particles following the Lagrangian dynamical system from some initial distribution.

The evolution equation of $\mP_L(\vPsi; \tau)$ could then be derived simply by taking the 
time derivatives of equation (\ref{eqn:PL_rel}), 
\begin{eqnarray}
\label{eqn:PL_transport_eq_deriv}
\frac{\partial}{\partial \tau} \mP_L (\vPsi; \tau) &=& 
\left \langle  \frac{\partial}{\partial \tau} \mP^f_L (\vPsi; \tau) \right \rangle_L 
=\left\langle \frac{d \psi_{\alpha}}{d \tau} 
\left[ \frac{\partial}{\partial \psi_{\alpha}}
\delta_D( \vpsi(\tau) - \vPsi) \right] \right\rangle_L  \nonumber \\
&=& - \left \langle  \chi_{\alpha}  \left [\frac{\partial}
{\partial \Psi_{\alpha} } P^f_L(\vPsi; \tau)  \right] \right \rangle_L
= - \frac{\partial}{\partial \Psi_{\alpha}} \left\langle  
\chi_{\alpha}  \mP^f_L(\vPsi;\tau)   \right \rangle_L. 
\end{eqnarray}
Here, we have changed the derivative variable from $\psi_{\alpha}$ to $\Psi_{\alpha}$, 
since $ \partial \delta_D(\vpsi-\vPsi)/\partial \psi_{\alpha} = - \partial \delta_D(\vpsi-\vPsi)/\partial \Psi_{\alpha} $. 
And we have also substituted the dynamical equation $d \psi_{\alpha}/ d\tau = \chi_{\alpha}$. 
In the last equality, the partial derivative with respect to the sample space variable 
$\partial / \partial \Psi_{\alpha}$ has been taken out of the average, since it
commutes with both random variables as well as the average operation $\langle \cdot \rangle_L$. 
Before proceeding, we first notice that, the right hand side of equation (\ref{eqn:dyn_eqn_full}), 
$\vchi[\vpsi, \varepsilon_{ij}]$, depends not only on $\vpsi$ but also on another random variable, 
specifically the tidal tensor $\varepsilon_{ij}$. Therefore, the ensemble average 
$\langle \chi_{\alpha} \mP^f_L(\vPsi; \tau) \rangle_L$ should be performed regarding the joint 
PDF $\mP_L(\vPsi, \vChi; \tau)$, where $\vChi$ is the sample space variable corresponding to $\vchi$. 
This leads to
\begin{eqnarray}
\langle \chi_{\alpha} \mP^f_L(\vPsi; \tau) \rangle_L 
&=& \int d\vPsi^{\pri} d\vChi^{\pri} ~  X^{\pri}_{\alpha}  \delta_D(\vPsi^{\pri}-\vPsi)
\mP_L(\vPsi^{\pri}, \vChi^{\pri}; \tau)   \nonumber \\
&=& \int d\vChi^{\pri} ~ X^{\pri}_{\alpha}  \mP_L(\vChi^{\pri} | \vPsi; \tau) \mP_L(\vPsi; \tau )
= \left\langle \chi_{\alpha} | \vPsi; \tau \right \rangle_L  \mP_L(\vPsi;\tau) , 
\end{eqnarray}
where we have expressed the joint PDF by the product of $\mP_L(\vPsi; \tau)$ and the 
conditional PDF $\mP_L(\vChi | \vPsi; \tau) = \mP_L(\vChi | \vpsi=\vPsi; \tau)$. 
Finally, we show that the evolution of $\mP_L(\vPsi; \tau)$ is simply described by the conservation
equation in multidimensional parameter space $\vPsi$ 
\begin{eqnarray}
\label{eqn:LPDF_evol_eqn}
\frac{\partial}{\partial \tau} \mP_L (\vPsi; \tau) + \frac{\partial}{\partial \Psi_{\alpha}}  
\biggl [  \left\langle \chi_{\alpha} | \vPsi; \tau \right \rangle_L  \mP_L(\vPsi;\tau) \biggr ] = 0, 
\end{eqnarray}
with the convective coefficients characterised by the conditional average
$\left \langle \vchi | \vPsi;  \tau \right \rangle_L $.

\subsection{Back to the Dynamics}

In the following, we will demonstrate that the evolution equation (\ref{eqn:LPDF_evol_eqn}) encodes 
enough dynamical information about the Lagrangian evolution of  fluid elements. 
Intuitively, this is comprehensible because the only reason that could distort the shape of 
PDF in the parameter space of $\vPsi$ is because there are genuine particles moved to that region. 
As a linear partial differential equation (PDE), one of the standard analytical methods to solve this equation
is the method of characteristics. 
With this method, the solution to equation (\ref{eqn:LPDF_evol_eqn}) is represented as the surface 
fabricated by the union of characteristic curves, 
which are defined as the integral curves of the vector field determined by the coefficients of the PDE. 
For our purpose, however, instead of the full solution, we are mostly interested in its projection to the 
dynamical variable space of $\vPsi$. 
Following the standard procedure, the projected characteristic trajectory is expressed as
\begin{eqnarray}
\label{eqn:effective_traj_Lag}
\frac{d}{d\tau}\vPsi (\tau) = \langle \vchi |\vPsi ; \tau \rangle_L. 
\end{eqnarray}
This ordinary differential equation resembles the full dynamical system  (\ref{eqn:dyn_sys_tot})
except for the additional operation of average conditional on the value of dynamical variables 
$\vPsi$ at time $\tau$.
Therefore, it would eliminate the dependence on extra degree of freedom other than $\vPsi$, 
and the trajectory then becomes localized.  
Since the only extra term in $\vchi$ is the tidal tensor $\varepsilon_{ij}$, after the conditional average, 
there will be no modification needed for most part of equation (\ref{eqn:dyn_eqn_full}) except for the
evolution of velocity shear
\begin{eqnarray}
\label{eqn:closed_dyneq}
\frac{d}{d\tau} \Sigma_{ij}  + \mH(\tau) \Sigma_{ij} + \frac{2}{3}\Theta\Sigma_{ij}
+ \Sigma_{ik}\Sigma^k_j  - \frac{1}{3}\Sigma_{mn}\Sigma^{mn}  \delta^K_{ij} =  
 - \langle \varepsilon_{ij} | \vPsi ; \tau \rangle_L . 
\end{eqnarray}
Therefore, at any given time $\tau$, our solution is obtained by averaging over  
environmental tidal effects among all fluid elements with the same value of density, velocity 
divergence and velocity shear. 
So this procedure would produce a mean effective fluid element 
whose trajectory in $\vpsi$ space  is integrated along the conditional averaged change rate of shear 
tensor.
Here we do want to emphasize that, one of the reason this method works is because
the coefficient in front of $\partial \mP /\partial \tau$ is unity, so that the parameter that characterizes
the integral curve is the time itself.

Interestingly, if one takes the equation (\ref{eqn:effective_traj_Lag}) as the original system, and repeats 
procedures from the last section, one would get the exact same $\mP_L$ evolution equation 
(\ref{eqn:LPDF_evol_eqn}). 
That is to say, although the dynamical behavior of a genuine fluid element is very different from 
the effective particles that described by equation  ({\ref{eqn:effective_traj_Lag}}), their one-point 
statistical will always be identical, provided that the coefficient
$\langle \vchi | \vPsi ;\tau \rangle_L$ is correctly modeled or measured. 
Therefore, if the distribution of a collection of effective particle in $\vpsi$ space is the same as 
the real fluid particles, their Lagrangian probability density function $\mP_L$ will remain the same
all the time \footnote{This is true as long as fluid description in (\ref{eqn:dyn_eqn_full}) is valid, 
which of course does not include the situation after the shell-crossing for dark matter particles.}. 
In the following, we will describe this as the statistical equivalence of effective fluid particles.

In the right panel of Figure. (\ref{fig:PDF_illustration}), we schematically illustrate the procedure
for obtaining this solution from the evolution equation of PDF. 
The surface represents the time evolution of the probability distribution, from a Gaussian to
some skewed distribution. The characteristic curve, shown as dash line, is then projected to the 
$\vPsi-\tau$ plane. 
With given initial condition $\vpsi(\tau_i)$, unlike genuine dynamical trajectories that resides 
somewhere within the grey band, the projected characteristic curve is uniquely determined by 
the initial condition.

\subsection{Lagrangian Evolution from the Eulerian Perspective }

To proceed, we have to estimate the Lagrangian conditional average $\langle \vchi | \vPsi ;\tau \rangle_L$, 
either by numerical measurement or analytical calculation. 
For the former, the Lagrangian average could be naturally carried out by sampling all particles in a N-body 
simulation. For analytical estimation, however, it will be more convenient to work in the Eulerian space
for calculating the gravitational potential $\Phi$ and tidal tensor $\varepsilon_{ij}$. 
But simply replacing previous derivation with the Eulerian probability density function ($\mP_E$)
would not work.  This is because the cosmic flow is highly compressible, $\mP_E$ 
is in general not the same as the Lagrangian counterpart $\mP_L$, nor does its evolution described
by equation (\ref{eqn:LPDF_evol_eqn}). 
Nevertheless, since the particle based  $\mP_L$ is equivalent to the density-weighted Eulerian 
probability density function ($\rho$PDF) with almost vanishing initial density perturbation $\drho(\tau_i)
\approx 0$, one could then proceed by defining \citep{Pop85} this quantity as
\begin{eqnarray}
\mD (\vPsi;\tau)  = (1+\Drho) \mP_E(\vPsi;\tau). 
\end{eqnarray}

Despite its conceptual straightforwardness, it is more complicated to formally obtain the 
evolution equation for $\mD(\vPsi;\tau)$. Therefore, we present the full derivation in Appendix A. 
As it turned out, for statistical homogeneous and isotropic field,  and assuming $\mD$ is bounded 
in velocity space, we recover a very similar evolution equation
\begin{eqnarray}
\frac{\partial}{\partial \tau}\mD(\vPsi;\tau) + \frac{\partial}{\partial \Psi_{\alpha}} \biggl [ 
\langle \chi_{\alpha} |\vPsi; \tau \rangle_E \mD(\vPsi;\tau) \biggr ]  = 0.
\end{eqnarray}
with the only apparent difference being the Eulerian conditional average 
$\langle \chi_{\alpha} |\vPsi; \tau \rangle_E$ instead of Lagrangian average. 
However, if we identify $\mP_L =  (1+\Drho) \mP_E = \mD$, then by definition 
\begin{eqnarray}
\langle \chi_{\alpha} | \vPsi ;\tau \rangle_L &=& \frac{1}{(1+\Drho)\mP_E(\vPsi) } \int d\vChi  ~
\Chi_{\alpha} (1+\Drho)\mP_E( \vChi, \vPsi) 
= \langle \chi_{\alpha} | \vPsi; \tau \rangle_E .
\end{eqnarray}
Therefore, in the following, we will simply neglect all subscripts $E/L$ as they are identical in
conditional average. 
And the mean evolution of given fluid particle will again be defined as the projected 
characteristic curve
\begin{eqnarray}
\label{eqn:effective_traj_E}
\frac{d}{d\tau}\Psi_{\alpha}(\tau) = \langle \chi_{\alpha} |\vPsi; \tau \rangle. 
\end{eqnarray}
In this context, the statistical equivalence of these trajectories is then stated regarding the
density-weighted PDF $\mD(\vPsi; \tau)$ in the Eulerian space.

\section{Statistical Closure of the Tidal Tensor}

\subsection{Gaussian Closure}

Since the early stage dynamical evolution, described by the Zel'dovich approximation, 
is local \citep{HB96}, to be a self-consistent localization procedure, one expects the statistical closure
method would recover the same dynamics with e.g. the Gaussian initial condition.
In the following, we will demonstrate this is indeed the case. 
Since the tidal tensor $\varepsilon_{ij}$ could be expressed either as the second order gradient or
the integration of density field, one has the freedom to perform the calculation in either formalisms.   
For Gaussian distributed field, the derivations are somewhat equally straightforward, 
as one need to deal with tensor correlation function for the first and the spatial integration of correlation
function for the latter. 
However, the local format does benefit the higher order calculation as the integration of higher order
correlation function becomes less appealing. 
On the other hand, the integral form provides a clearer picture of how our method solve the
non-local problem. 
So we will present the derivation of the Gaussian closure in the integral form in Appendix 
\ref{appsec:gaussclos_intg}.

Writing in the local derivative form, the definition of conditional average is expressed as the integration 
of the joint probability function $\mP(\vE, \vPsi; \tau) = \mP(\vGamma;\tau)  $, 
\begin{eqnarray}
\label{eqn:eij_cond_avg_def}
\langle \varepsilon_{ij} | \vPsi ; \tau \rangle \mP(\vPsi; \tau)
= \int d\vE ~ E_{ij} \mP(\vE, \vPsi; \tau)  =  \int d\vE ~ E_{ij} \mP(\vGamma; \tau) 
\end{eqnarray}
Here we have defined the new
variable $\vGamma = \{ \vE,  \vPsi \}$, where $\vE=\{E_{ij}\}$ is the sample space variable of
peculiar tidal tensor $\varepsilon_{ij}$. 
In Appendix \ref{appsec:gaussian}, we present the detailed derivation of the conditional average
assuming Gaussian distributed variables. The result is quite obvious for experienced cosmologist,
but since we will also extend the calculation to non-Gaussian field, we would like to briefly 
highlight the procedure here.
Among others, one critical point that enables the calculation is to introduce the characteristic function
or partition function, which is defined as the Fourier transform of the PDF
\begin{eqnarray}
\mZ(\vlambda; \tau)  = \int d\vGamma  ~ e^{i \vlambda \cdot \vGamma} ~ \mP(\vGamma; \tau). 
\end{eqnarray}
So the Gaussian partition function is simply $\mZ_G(\vlambda) = \exp(- \lambda_{\alpha} 
\xi_{\alpha\beta} \lambda_{\beta} /2 )$, where $\xi_{\alpha\beta}$ is the component of covariance matrix. 
By expressing $\mP(\vGamma)$ as the inverse Fourier transform of $\mZ(\vlambda)$ at the right hand
side of equation (\ref{eqn:eij_cond_avg_def}), one would then 
be able to replace $E_{ij}$ with derivatives. 
After a few steps of such manipulations, eventually one would obtain 
(equatoin \ref{eqn:cond_avg_gauss_derv})
\begin{eqnarray}
\langle \varepsilon_{ij} | \vPsi; \tau \rangle \mP(\vPsi; \tau)  &=& - \xi^{\varepsilon\psi}_{ij, \alpha} 
\left (  \frac{\partial}{\partial  \Psi_{\alpha}}  \mP(\vPsi) \right). 
\end{eqnarray}
where $\xi^{\varepsilon \psi}_{ij,\alpha}$ is the covariance matrix between tidal tensor $\varepsilon_{ij}$
and dynamical variable $\psi_{\alpha}$. 
Unlike the derivation in Appendix \ref{eqn:cond_avg}, we will not explicitly keep the order of the cumulant
matrix in the main text since it could be easily inferred from the number of the superscript variables or of 
indices. Here the tensor Latin indices pair ${ij}$ is interpreted as a single spatial index. We will use a 
comma to divide among such indices, please keep in mind that they are not derivatives, which
we will always write explicitly in this paper. 
Eventually, after substituting the definition of Gaussian $\mP(\vPsi)$,  the final result  is simply 
linearly proportional to $\vPsi$ (equation \ref{eqn:cond_avg_gauss}), 
\begin{eqnarray}
\langle \varepsilon_{ij} | \vPsi ; \tau \rangle =
\xi^{\varepsilon \psi}_{ij,\alpha} \left( \xi^{\psi\psi} \right)^{-1}_{\alpha\beta} \Psi_{\beta}, 
\end{eqnarray}
where $\left( \xi^{\psi\psi}\right )^{-1}_{\alpha\beta}$ is the component of the inverse covariance 
matrix between $\vpsi$. 

Therefore, the calculation will be straightforward as long as the covariance matrix between 
$\varepsilon_{ij}$ and $\vPsi$ is given. 
In Appendix \ref{appsec:covmat}, we present all relevant covariance matrices and the 
inverse.  As shown there, the only non-vanishing component of $\xi^{\varepsilon\psi}$ is 
$\xi^{\varepsilon A}_{ij,mn}$, therefore  
\begin{eqnarray}
\langle \varepsilon_{ij} | \vPsi ; \tau \rangle & = & \xi^{\varepsilon A}_{ij,mn} \left[  
\left(\xi^{-1}_{\psi\psi}\right )^{A\delta}_{mn} \Drho +  
\left(\xi^{-1}_{\psi\psi}\right )^{A A}_{mn, kl} \mA_{kl} \right] .
\end{eqnarray}
After substituting all components of these matrix, and then contracting indices of Kronecker delta functions,  
the first term that proportional to $\Drho$ vanishes, with the only contribution 
\begin{eqnarray}
\label{eqn:eij_gauss_clos}
\langle \varepsilon_{ij} | \vPsi ; \tau \rangle & = & \frac{4\pi G_N \bar{\rho} a^2}{45}
\left( \frac{15  \sigma^2_{\delta\theta} }{2 \sigma^2_{\theta\theta} }  \right) 
 \left (   3\delta^K_{im}\delta^K_{jn} + 3\delta^K_{in}\delta^K_{jm}
  - 2\delta^K_{ij}\delta^K_{mn} \right) \mA_{mn} \nonumber\\
   & = & 4\pi G_N \bar{\rho} a^2 \left( \frac{ \sigma^2_{\delta\theta}   }{ \sigma^2_{\theta\theta} } \right)
 \left( \mA_{ij} - \frac{\Theta}{3}  \delta^K_{ij}  \right)  
 =  4\pi G_N \bar{\rho} a^2  \left( \frac{ \sigma^2_{\delta\theta}   }{ \sigma^2_{\theta\theta} } \right) \Sigma_{ij}.
\end{eqnarray}
Therefore, the conditional averaged tidal effect on a fluid element embedded in a Gaussian random field 
is simply proportional to its velocity shear tensor $\sigma_{ij}$. 
Notice that we have assumed the velocity gradient tensor $A_{ij}$ is symmetric, therefore excluding the 
presence of vorticity. 
Furthermore, in the linear order, the continuity equation becomes
\begin{eqnarray}
\label{eqn:lin_cont}
\theta = -\mH(\tau) f(\tau) \drho, 
\end{eqnarray}
where $f(\tau) = d\ln D(\tau)/d\ln a$, and $D(\tau)$ is the linear growth factor. 
Hence the coefficient $\sigma^2_{\delta\theta}/\sigma^2_{\theta\theta} = -1/\mH(\tau)f(\tau)$. 
Consequently, our averaged tidal effect of a fluid element in a Gaussian density field coincides
with the prediction from Zel'dovich approximation \citep{HB96}
\begin{eqnarray}
\label{eqn:epsilon_za}
\varepsilon_{ij}= \frac{-4\pi G_N\bar{\rho} a^2  }{\mH f} \sigma_{ij}. 
\end{eqnarray}
And the dynamical system will be exactly the same as ZA if we further replace $\drho$  on the right hand 
side of Raychaudhuri equation (\ref{eqn:closed_dyneq})  with $\theta$ using this relationship 
\citep{HB96}, i.e.
\begin{eqnarray}
\frac{d}{d\tau} \theta + \mH(\tau)\theta + \frac{1}{3}\theta^2 + 
\sigma^{ij} \sigma_{ij}   = \frac{4\pi G_N \bar{\rho} a^2}{\mH f} \theta. ~
\end{eqnarray}

We would like to emphasize that, despite being exact the same as ZA prediction, our result here is 
nontrivial and fundamentally different. Instead of solving linearized dynamics, our dynamical system
presents the effectively localized trajectories that preserve the density weighted probability 
density function in a Gaussian random density field. 
The coincidence here, however, is actually demanded since the ZA itself is local. 
So in the early stage of structure formation where ZA applies, the localized effective trajectory
is just ZA itself.
From this point of view, the result in equation (\ref{eqn:eij_gauss_clos}) demonstrates the self-consistency 
of our method. 
And the way this was achieved highlights the intriguing connection between linearized dynamics and 
Gaussian  statistics. 

Moreover, it is well known that the ZA evolution would provide an incorrect solution for spherical infall, 
because it does not obey the Poisson equation. 
In our formalism, this is not necessarily the case since we did not make any simplification 
until equation (\ref{eqn:lin_cont}) and (\ref{eqn:epsilon_za}). 
Without doing so, by setting  $\varepsilon_{ij}=\sigma_{ij}=0$ in equation 
(\ref{eqn:closed_dyneq}), one would recover spherical collapse (SC) model exactly. 
Therefore, our Gaussian closure here is consistent with both ZA and spherical collapse, 
while the first is obtained in the limit of linearized continuity equation (\ref{eqn:lin_cont}) 
and the latter for certain geometry of the fluid element.

One caveat of our derivation is that, by assuming the linear continuity equation (\ref{eqn:lin_cont}), 
the variance relation $\sigma^4_{\delta\theta} - \sigma^2_{\delta\delta}\sigma^2_{\theta\theta} =0$. 
As shown from Appendix \ref{appsec:covmat}, the inverse matrix of 
$\xi^{-1}_{\psi\psi}$ becomes singular. Fortunately, it was the only non-singular term ($D_3$ term) that
eventually entered our calculation.
A conceptually more rigorous procedure is to consider a tiny non-linear density $\drho$ and velocity 
divergence $\theta$, e.g. up to the second order, and obtain the same result by taking the limit where 
the non-linearity approaches to zero.

\subsection{Mean Tidal Tensor in the Weakly Non-Gaussian Field}

One advantage of our approach is that by re-formulating the non-local gravitational field theory into 
a set of ordinary differential equations, no other approximations (e.g. perturbative expansion of 
dynamical variables)  has been made sacrifying the non-linearity of the dynamical system. 
Of course, this is under the assumption that the conditional average of the tidal tensor is estimated 
in non-linear  regime as well. 
While it is obviously complicated to evaluate in the deeply non-linear regime, we will first utilize 
the cumulant expansion theorem \citep{Ma85,Mat03} to calculate the corrections to the next  
order, i.e. up to the third-order cumulants. 
The cumulant expansion theorem states that the logarithm of the partition function could be expanded
by $n$-th order of cumulants
\begin{eqnarray}
\ln \mZ(\vlambda) = \sum_{n\ge 1} \left( \frac{i^n}{n!} \right) \xi^{(n)}_{\alpha_1\cdots\alpha_n} 
~ \lambda_{\alpha_1} \cdots \lambda_{\alpha_n}. 
\end{eqnarray}
Transforming back to probability density function and substituting the current with partial derivative 
$\lambda_{\alpha} = i \partial /\partial \Gamma_{\alpha}$, one obtains the expansion of arbitrary 
probability density function $\mP(\vGamma)$ in terms of  Gaussian distribution $\mP_G(\vGamma)$
\citep{Cra1946,Ken1958,Mat1994,Mat03,PGP09}
\begin{eqnarray}
\mP(\vGamma) = \exp\left [  \sum_{n\ge 3} \frac{(-1)^n}{n!} \xi^{(n)}_{\alpha_1\cdots \alpha_n} 
\frac{\partial ^n }{ \partial \Gamma_{\alpha_1} \cdots \partial \Gamma_{\alpha_n}}  \right] \mP_G(\vGamma).
\end{eqnarray}
From the definition of the conditional average of tidal tensor in equation (\ref{eqn:eij_cond_avg_def}), one
then has to expand both $\mP(\vGamma)$ as well as $\mP(\vPsi)$ at both sides of the equality.
Then the derivation is quite straightforward though a little tedious, and we will present all the details 
in Appendix \ref{appsec:nongaussian}.

From equation (\ref{eqn:cond_avg_skew}), the conditional averaged $\varepsilon_{ij}$ could be
expressed as the Gaussian result $ \langle \varepsilon_{ij}| \vPsi \rangle_G$ plus a correction 
term that is related to the second and third order cumulants matrices 
\begin{eqnarray}
\langle \varepsilon_{ij}| \vPsi \rangle  &=& \langle \varepsilon_{ij}| \vPsi \rangle_G + 
\frac{1}{2} \left [ \xi^{\varepsilon\psi\psi}_{ij,\alpha\beta} -  \xi^{\psi\psi\psi}_{\alpha\beta\gamma}  
 \xi^{\varepsilon\psi}_{ij, \delta} (\xi^{\psi\psi})^{-1}_{\delta\gamma} \right] 
 \left [ - \left (  \xi^{\psi\psi}  \right)^{-1}_{\alpha\beta} + \left (\xi^{\psi\psi} \right )^{-1}_{\alpha\lambda}
 \left( \xi^{\psi\psi} \right)^{-1}_{\beta\tau} \Psi_{\lambda} \Psi_{\tau} \right ]. 
\end{eqnarray}
First of all, we notice from Appendix \ref{appsec:nongaussian} that the contribution 
$ \xi^{\varepsilon\psi}_{ij, \delta} (\xi^{\psi\psi})^{-1}_{\delta\gamma} $ 
arises from the derivative $\partial /\partial \Psi_{\gamma} \langle \varepsilon_{ij} |\vPsi \rangle_G$. 
And since the Gaussian $ \langle \varepsilon_{ij} |\vPsi \rangle_G$  is simply proportional to 
$\Sigma_{ij} = (\delta^K_{im}\delta^K_{jn} - \frac{1}{3}\delta^K_{mn}\delta^K_{ij} ) \mA_{mn} $, 
this term then reduces to
\begin{eqnarray}
 \xi^{\psi\psi\psi}_{\alpha\beta\gamma}  \xi^{\varepsilon\psi}_{ij, \delta} (\xi^{\psi\psi})^{-1}_{\delta\gamma}
 & =&  \xi^{A\psi\psi}_{mn,\beta\gamma}  
  (4\pi G_N \bar{\rho} a^2) \left( \frac{\sigma^2_{\delta\theta}}{\sigma^2_{\theta\theta}}  \right) 
 \left ( \delta^K_{im}\delta^K_{jn} - \frac{1}{3}\delta^K_{mn}\delta^K_{ij} \right) \nonumber\\
 &=& (4\pi G_N \bar{\rho} a^2) \left( \frac{\sigma^2_{\delta\theta}}{\sigma^2_{\theta\theta}}  \right) 
\xi^{\sigma\psi\psi}_{ij,\alpha\beta}. 
\end{eqnarray}
From Appendix \ref{appsec:cskewmat}, we show that the third order cumulant matrix 
$\xi^{\varepsilon\psi\psi}_{ij,\alpha\beta}$ and $\xi^{\sigma\psi\psi}_{ij,\alpha\beta}$ are quite similar, 
with only slightly different coefficients. 
Therefore, it will be very convenient to define the combined three-order cumulant matrix 
\begin{eqnarray}
\xi^{(\varepsilon - \sigma)\psi\psi}_{ij,\alpha\beta} &=& \xi^{\varepsilon \psi\psi}_{ij,\alpha\beta} -
(4\pi G_N \bar{\rho} a^2) \left( \frac{\sigma^2_{\delta\theta}}{\sigma^2_{\theta\theta}}  \right) 
\xi^{\sigma\psi\psi}_{ij,\alpha\beta} \nonumber \\
&=&  (4\pi G_N \bar{\rho} a^2)  \int \frac{d\vk_{123}}{(2\pi)^6} \delta_D(\vk_{123})
\left( \hk_{1i} \hk_{1j} - \frac{1}{3}\delta^K_{ij} \right) \hk_{2m}\hk_{2n}  
 \hk_{3k}\hk_{3l} ~ B^{(\delta-\theta)\theta\theta}(\vk_1, \vk_2, \vk_3), 
\end{eqnarray}
where the combined bispectrum is defined similarly
\begin{eqnarray}
 B^{(\delta-\theta)\theta\theta}(\vk_1, \vk_2, \vk_3)
 =  B^{\delta\theta\theta}(\vk_1, \vk_2, \vk_3) - 
 (4\pi G_N \bar{\rho} a^2) \left( \frac{\sigma^2_{\delta\theta}}{\sigma^2_{\theta\theta}}  \right) 
  B^{\theta\theta\theta}(\vk_1, \vk_2, \vk_3). 
\end{eqnarray}

Now let us consider the contribution $\xi^{(\varepsilon - \sigma)\psi\psi}_{ij,\alpha\beta} 
\left( \xi^{\psi\psi} \right)^{-1}_{\alpha\beta} $. Explicitly expanding all summations, 
\begin{eqnarray}
\label{eqn:vanish_contr}
\xi^{(\varepsilon - \sigma)\psi\psi}_{ij,\alpha\beta}  \left( \xi^{\psi\psi} \right)^{-1}_{\alpha\beta} 
 = \xi^{(\varepsilon - \sigma)\delta\delta}_{ij} \left( \xi^{-1} \right)^{\delta\delta}
 +  2  \xi^{(\varepsilon - \sigma)\delta A}_{ij,mn} \left( \xi^{-1} \right)^{\delta A}_{mn}
 +   \xi^{(\varepsilon - \sigma)A A}_{ij,mn, kl} \left( \xi^{-1} \right)^{A A}_{mn,kl}. 
\end{eqnarray}
As shown from Appendix \ref{appsec:cskewmat}, 
the matrix component $\xi^{(\varepsilon - \sigma)\delta\delta}_{ij}  $ is simply zero. 
For the matrix component $\xi^{(\varepsilon - \sigma)\delta A}_{ij, mn} $, it is proportional to 
$ 3 \delta^K_{im}\delta^K_{jn} + 3\delta^K_{in} \delta^K_{jm} - 2 \delta^K_{ij} \delta^K_{mn} $, 
whereas $ \left( \xi^{-1} \right)^{\delta A}_{mn} \sim \delta^K_{mn} $. Therefore, their
contraction also vanishes. 
Finally, the third contribution in equation (\ref{eqn:vanish_contr}) reads
\begin{eqnarray}
\xi^{(\varepsilon - \sigma)A A}_{ij,mn,kl}  \left( \xi^{-1} \right)^{A A}_{mn,kl} 
\sim \int \frac{d\vk_{123}}{(2\pi)^6}
\delta_D(\vk_{123}) \left( \hk_{1i} \hk_{1j} - \frac{1}{3}\delta^K_{ij} \right) 
\left( D_3 \mu_{23}^2 + D_4 \right)   B^{(\delta-\theta)\theta\theta}(\vk_1, \vk_2, \vk_3), 
\end{eqnarray}
where coefficients $D_3$ and $D_4$ are defined in equation (\ref{eqn:xipsipsi_inv}), and we
have denoted $\vk_{123}=\vk_1+\vk_2+\vk_3$. 
On the other hand, due to rotational invariance, this term could only be proportional $\delta^K_{ij}$. 
Then the proportional coefficient also vanishes, as can be easily seen by contracting with 
$\delta^K_{ij}$. 
Therefore, the correction term to the conditional average is simply quadratic in $\vPsi$, 
and linearly proportional to the third order cumulant matrix
\begin{eqnarray}
\Delta_{\langle \varepsilon_{ij}| \vPsi \rangle} = 
\langle \varepsilon_{ij}| \vPsi \rangle  - \langle \varepsilon_{ij}| \vPsi \rangle_G 
= \frac{1}{2}
\xi^{(\varepsilon - \sigma)\psi\psi}_{ij,\alpha\beta} (\xi^{\psi\psi})^{-1}_{\alpha\delta}  
~(\xi^{\psi\psi})^{-1}_{\beta\lambda} \Psi_{\delta} \Psi_{\lambda}, 
\end{eqnarray}
After substituting all matrices elements and some simple algebra, we could eventually write the correction 
terms up to $\xi^{(3)}$ as
\begin{eqnarray}
\Delta_{\langle \varepsilon_{ij}| \vPsi \rangle} = 
 \left(Q_{\rho} \Drho +  Q_{\theta} \Theta \right)   \Sigma_{ij}
 +  Q_{\Sigma^2}  (\widetilde{\Sigma^2})_{ij} , 
\end{eqnarray}
where $Q_{\rho}, Q_{\theta}$ and $Q_{\Sigma^2}$ are coefficients, 
and $(\widetilde{\Sigma^2})_{ij}$ is the traceless part of $(\Sigma^2)_{ij}$
\begin{eqnarray}
(\widetilde{\Sigma^2})_{ij} =  \Sigma_{i}^{~m}\Sigma_{mj} - \frac{1}{3} 
(\Sigma^{mn}\Sigma_{mn})  \delta^K_{ij} . 
\end{eqnarray}
With the definition of various two- and three-point correlations shown in Appendix 
\ref{appsec:cumulant_matrix}, the coefficients could be expressed as
\begin{eqnarray}
Q_{\rho}& = &  D_2 D_3  \left [ \frac{1}{5}  \tilde{\xi}^{(\varepsilon-\sigma) AA}_1 + 
   6 \left( \frac{D_1}{D_2} \right)  \xi^{(\varepsilon-\sigma) \delta A} \right] 
   = D_2 D_3   \xi_{\rho}^{(\varepsilon-\sigma) (A-\delta) A}    \nonumber\\
Q_{\theta} &=& D_3 D_5 \left [  \frac{1}{5}   \tilde{\xi}^{(\varepsilon-\sigma) AA}_1 + 
   6 \left( \frac{D_2}{D_5}\right)  \xi^{(\varepsilon-\sigma) \delta A} \right]  
      = D_3 D_5   \xi_{\theta}^{(\varepsilon-\sigma) (A-\delta) A}    \nonumber \\
Q_{\Sigma^2}  &=& 4 D_3^2 \xi^{(\varepsilon - \sigma) A A}_4. 
\end{eqnarray} 
Note that we have defined two new quantities $ \xi_{a}^{(\varepsilon-\sigma) (A-\delta) A}$  for 
$Q_{\rho}$ and $Q_{\theta}$ respectively, where $a=\rho$ or $\theta$. 
In Fourier space, 
\begin{eqnarray}
  \xi_{a}^{(\varepsilon-\sigma) (A-\delta) A}
    = \left ( 4\pi G_N \bar{\rho} a^2 \right) \int \frac{d\vk_{123}} {(2\pi)^6} \delta_D(\vk_{123})
    \left ( \frac{3 \mu^2_{13 } -1}{15} \right) B_{a}^{(\delta-\theta)(\theta-\delta)\theta} (\vk_1,\vk_2,\vk_3), 
\end{eqnarray}
where we have further defined 
\begin{eqnarray}
    B_{\rho}^{(\delta-\theta)(\theta-\delta)\theta}&=& 
 B^{(\delta-\theta)\theta\theta} - \left( 
    \frac{\sigma^2_{\theta\theta}}{\sigma^2_{\delta\theta}} \right) B^{(\delta-\theta)\delta\theta } 
    =  B^{\delta\theta\theta} - \left( \frac{\sigma^2_{\delta\theta}}{\sigma^2_{\theta\theta}} \right) 
    B^{\theta\theta\theta} -  \left( \frac{\sigma^2_{\theta\theta}}{\sigma^2_{\delta\theta}} \right) 
    B^{\delta\delta\theta} + B^{\theta\delta\theta}   \nonumber\\
 B_{\theta}^{(\delta-\theta)(\theta-\delta)\theta}&=& 
 B^{(\delta-\theta)\theta\theta} - \left( 
    \frac{\sigma^2_{\delta\theta}}{\sigma^2_{\delta\delta}} \right) B^{(\delta-\theta)\delta\theta } 
    =  B^{\delta\theta\theta} - \left( \frac{\sigma^2_{\delta\theta}}{\sigma^2_{\theta\theta}} \right) 
    B^{\theta\theta\theta} -  \left( \frac{\sigma^2_{\delta\theta}}{\sigma^2_{\delta\delta}} \right) 
    B^{\delta\delta\theta} + \left( \frac{ \sigma^4_{\delta\theta} }{ \sigma^2_{\delta\delta} 
    \sigma^2_{\theta\theta} } \right) B^{\theta\delta\theta}  . \nonumber \\
\end{eqnarray}
At the tree level, $\sigma^2_{\theta\theta} /\sigma^2_{\delta\theta}= 
\sigma^2_{\delta\theta} / \sigma^2_{\delta\delta} =-\mH f$, 
and the bispectrum $B^{abc}(\vk_1,\vk_2,\vk_3)$ is simply related to linear power spectrum $P_L$ as
\begin{eqnarray}
B^{abc}(\vk_1,\vk_2,\vk_3)  &=& (- \mH f)^n 2 \biggl[ K^{(2)}_a(\vk_2, \vk_3) P_L(k_2)  P_L(k_3) +   
K^{(2)}_b(\vk_1 \vk_3) P_L(k_1)  P_L(k_3)   \nonumber\\
&&  +  K^{(2)}_c(\vk_1, \vk_2) P_L(k_1)  P_L(k_2) \biggr ],
\end{eqnarray}
where $a, b, c=\{ 0, 1\} = \{\delta, \theta\}$, and $n=a+b+c$. 
We have also redefined the second order perturbation kernel as $K^{(2)} = \{ F^{(2)} , G^{(2)} \}$. 
It is straightforward to check that both $ B_{\rho,\theta}^{(\delta-\theta)(\theta-\delta)\theta}$ vanish
at this order. However, this does not mean the coefficients $Q_{\rho}$ and  $Q_{\theta}$ would be zero, since
$Q_2$ and $Q_5$ also diverge in this limit. 
Hence one has to evaluate to the one-loop order. To proceed, we define the
correction to the Gaussian variance  
\begin{eqnarray}
\sigma^2_{ab} =  \sigma^2_{ab,G} ( 1 + \gamma_{ab} ),  \qquad a,b = \{ \delta, \theta\}, 
\end{eqnarray}
where $ \sigma^2_{ab,G} $ is the Gaussian variance. 
So the combined bispectrum could be decomposed into two terms
\begin{eqnarray}
 B_{\rho}^{(\delta-\theta)(\theta-\delta)\theta}  = T_1 + T_2 = \left [
 B^{\delta\theta\theta} + \left( \frac{1}{\mH f} \right)
    B^{\theta\theta\theta} +  (\mH f)   B^{\delta\delta\theta} + B^{\theta\delta\theta}  \right]
+ \frac{ \gamma_{\delta\theta} - \gamma_{\theta\theta} }{\mH f} \left[ B^{\theta\theta\theta} - (\mH f)^2 
B^{\delta\delta\theta} \right], 
\end{eqnarray}
where bispectra need to be evaluated up to the one-loop order in term $T_1$, but only to the tree-level 
in term $T_2$.  
Similarly, we have $ B_{\theta}^{(\delta-\theta)(\theta-\delta)\theta} = T_1 + T_3 $, and
\begin{eqnarray}
T_3 = (\gamma_{\delta\theta} - \gamma_{\theta\theta} ) \left [  B^{\theta\delta\theta} + 
\left ( \frac{1}{\mH f} \right)  B^{\theta\theta\theta}  \right] 
+ (\gamma_{\delta\theta} - \gamma_{\delta\delta} ) \left [  B^{\theta\delta\theta} + 
   (\mH f ) B^{\delta\delta\theta}  \right] . 
\end{eqnarray}

From above discussion, it is reasonable to expect that the conditional average of the tidal tensor could in 
general written as some function of dynamical variables $\vPsi$, i.e. 
\begin{eqnarray}
\langle \varepsilon_{ij} | \vPsi ; \tau \rangle =  \mathscr{F}_{ij} (\vPsi; \tau) . 
\end{eqnarray} 
It is clear that the function $\mathscr{F}_{ij}$ would be symmetric and traceless on spatial coordinates.  
And the coefficients of its Tylor series would depend on the statistics of the field.
Although we have denoted the function in a compact form $ \mathscr{F}_{ij}$, one should not assume its analyticity 
as it is not obvious to us the series would converge for arbitrary distribution of field $\vPsi$.

\section{Discussion and Conclusion}

\subsection{Stochasticity and Effective Fluid Elements}
So far, we have derived our effective evolution of a fluid element from the dynamical system described
by equations (\ref{eqn:Lagrangian_dyn_details}) - (\ref{eqn:Lagrangian_dyn_details_Aij}). 
As demonstrated in previous sections, the justification of the method arises from the dynamical information 
encoded in the PDF evolution equation (\ref{eqn:LPDF_evol_eqn}). 
In a Gaussian or weakly non-Gaussian field, it is also technically feasible as the statistical correlation between 
the tidal tensor $\varepsilon_{ij}$  and other dynamical variables $\vpsi$ is straightforward to estimate. 
Even in the deeply non-linear(Gaussian) region, one might still gain valuable information with inaccurate 
yet reasonable assumptions about their joint distribution, e.g. log-normal. 
However, in such regime, the fluid system itself becomes an incorrect approximation,
so even with contribution from tidal tensor, the set of our fluid variables $\vpsi$ would not suffice to 
describe  the system anymore. 
In the context of the CDM cosmology, the underlying distribution is about self-gravitating dark 
matter particles, which in general, are not necessarily the same as the `fluid parcels' described 
in this paper. Specifically, after the system entering into the multi-stream region, it is well known that 
equations  (\ref{eqn:Lagrangian_dyn_details}) - (\ref{eqn:Lagrangian_dyn_details_Aij}) would break down, 
and modifications should be implemented by introducing new contributions to these equations or 
adopting a more fundamental description of the system.
Neither of these approaches is currently well understood due to the difficulties of modeling the small 
scale phase space evolution, thus any alternatives would be welcome.

From the point of view of our statistical effective method, dynamical information will only be extracted 
from the one-point PDF evolution, regardless of particles' genuine trajectories.
This opens up a new perspective of approaching the problem. 
If the statistical information between these extra contributions and $\vpsi$ could be
known from some appropriate estimation of microscopic degree of freedoms, one could then introduce
a conditional average term similar to $\langle \varepsilon_{ij} | \vPsi; \tau \rangle$. 
On the other hand, a much cruder choice is just to phenomenologically introduce stochastic contributions 
that mimic the PDF evolution of the real system. 
Eventually, one would benefit from such simplification because of the inevitable randomness in the real world, 
even if it does not arise from effects one originally expected.
For examples, this could include the multi-streaming, complicated internal dynamics, and the arbitrary 
boundary shape of  finite-sized cosmic patch etc.

Especially, we are interested in the correction from the multi-streaming. 
Let us still work on the fluid system, after the shell-crossing, the evolution equation of velocity in 
Eulerian space would receive  contribution from the stress tensor. 
The standard procedure is to consider the one-point phase space density 
of CDM particles $f(\vx,\vp,\tau)$, with $\vp = a m \vu $ being the momentum, 
and its governing kinematic equation described by the collisionless 
Vlasov equation \citep{P80,BCGS02}. 
Then the continuity and Euler equation could be derived by simply taking the 
zeroth and first order momentum average respectively, with an extra contribution to the latter
\begin{eqnarray}
\label{eqn:zeta_def}
\varsigma_i(\vx) = - \frac{1}{\rho}\nabla_j (\rho\pi_{ij}).
\end{eqnarray} 
Here $\pi_{ij}$ is the second-order moment of phase space density $f(\vx,\vp,\tau)$, 
which is estimated by summing over all streams at given Eulerian position $\vx$.
The evolution of $\pi_{ij}$ could then be further derived by taking higher order moment and so on, 
which eventually produce a hierarchy of coupled equations. 
Even before worrying about the truncation of the hierarchy,  it is obvious that
after the shell-crossing, converting the quantity $\pi_{ij}$ or $\varsigma_i$ from 
Eulerian to Lagrangian space is conceptually problematic, since there is no well-defined 
notion of a single fluid element when collisionless dark matter particles simply 
bypass each other without much interaction. Clearly, this is an intrinsic defect of the 
Lagrangian fluid description in CDM scenario. 
Fortunately in our framework, it is the one-point PDF equation we are mostly 
interested in, and an Eulerian density-weighted PDF is always well-defined. 
Therefore, by generalizing the idea of mean evolution, we would like to consider 
some effective fluid particles, whose one-point $\rho$PDF will be identical to the 
real system.

We need to emphasize that even in Eulerian space, such contribution from a self-gravitational 
system is undoubtable not  purely random.
However, if one consider the relatively early stage of multi-streaming, or spatially some outer region of 
clusters, the number of streams is not necessarily large.  
To some extent (e.g. with finite time step), this contribution does carry certain stochastic traits.
In general, we would like to absorb any mean contribution into deterministic terms $\vchi$ so that 
the remaining part would have $\langle \zeta_{\alpha} ( \tau) \rangle = 0$. 
Furthermore, as the simplest model, we assume the process is Markovian, i.e. the 
correlation function  between time $\tau$ and $\tau^{\pri}$  
always vanishes except $\tau=\tau^{\pri}$,  
\begin{eqnarray}
\langle \zeta_{\alpha} (\tau) \zeta_{\beta}(\tau^{\pri}) \rangle = 
\xi^{\zeta}_{\alpha\beta} (\tau) \delta_D(\tau-\tau^{\pri}). 
\end{eqnarray}
As a result, our dynamical system then turns to
\begin{eqnarray}
\label{eqn:stochastic_diff_eq}
\frac{d}{d\tau} \psi_{\alpha} = \chi_{\alpha}[\vpsi,\varepsilon_{ij}] + \zeta_{\alpha}. 
\end{eqnarray}
And we are interested in the dynamical information encoded in the evolution equation of density 
weighted PDF $\mD(\vPsi;\tau)$, which becomes the Fokker-Planck equation
\begin{eqnarray}
\label{eqn:fokker_planck}
\frac{\partial}{\partial \tau}\mD(\vPsi; \tau) + \frac{\partial}{\partial \Psi_{\alpha}} \langle \chi_{\alpha} |\vPsi; \tau 
\rangle \mD(\vPsi; \tau)  = \frac{1}{2}
\xi^{\zeta}_{\alpha\beta}(\tau) \frac{\partial^2}{\partial \Psi_{\alpha} 
\partial \Psi_{\beta}} \mD (\vPsi; \tau).  
\end{eqnarray}
The solution of this equation is the Langevin equation. Combined with the result of previous
sections, we have
\begin{eqnarray}
\label{eqn:Langevin}
\frac{d}{d\tau}\Psi_{\alpha}(\tau) = \langle \chi_{\alpha} |\vPsi ; \tau\rangle + 
 \zeta_{\alpha}(\tau). 
\end{eqnarray}

One way to make use of equation (\ref{eqn:Langevin}) would be the Monte Carlo sampling
of dynamical trajectories in multi-streaming regime.  
Moreover, it could also provide interesting insights on physics at this scale. For example, the generation 
of vorticity.  As the velocity field is believed to be grown out of purely irrotational perturbation, the 
emergence of the vorticity could only be generated by the shell-crossing, or in fluid description 
from the contribution $\varsigma_i$ in equation (\ref{eqn:zeta_def}). 
Therefore, in the spirit of equation (\ref{eqn:Langevin}), one could interprete such generation
of rotational degree of freedom as the consequence of some stochastic 
process in the parameter space of velocity gradient tensor $A_{ij}$. 
So the evolution equation of vorticity $\omega_i$ will then be sourced by some stochastic 
terms
\begin{eqnarray}
\frac{d}{d\tau} \omega_i + \mH(\tau) \omega_i + \frac{2}{3} \theta \omega_i - 
\sigma_i^j \omega_j = (stochastic ~terms ). 
\end{eqnarray}
This coincides with the picture suggestion by \cite{WS14}. After defining the rotational invariants 
of tensor $A_{ij}$, which essentially combines both potential and rotational information of $A_{ij}$, 
\cite{WS14} demonstrated that the distinct correlation between vorticity generation and cosmic
web structure could be phenomenologically explained by a stochastic process that driving 
the transition from potential to rotational flow in their parameter space.

To some extent, this generalized dynamics has a similar underlying philosophy to the approach of the effective field theory (EFT) of the large-scale structure \citep{BEFT12,CHS12,PZ13,MP14}, where new operators that are compatible to the symmetries of the problem were introduced to effectively describe the evolution of large-scale modes. In principle, by comparing with the N-body simulations, the effects of smaller scales, including the shell-crossing regime, could be described by some low-energy constants. 
 With a very different motivation, here we are not particularly interested in the accuracy of large scale
 modes. Instead, since the effective dynamical equation (Eq. \ref{eqn:effective_traj_E}) starting from fluid dust model will certainly break down after the shell-crossing,  we argue that some effective terms, stochastic or not, should be introduced to preserve the statistic equivalence of the one-point PDF, which is the core of our method.

\subsection{Beyond the Mean Trajectory} 
For almost any statistical problem, the mean of an unknown distribution could only provide small amount of 
information. With given initial condition $\vPsi(\tau_i)$, we would also like to understand the
scattering around the mean trajectory, i.e. the conditional distribution 
\begin{eqnarray}
\mP \left (\vPsi | \vPsi_{\tau_i}; \tau \right) = 
\mP \left (\vPsi (\tau) | \vPsi(\tau_i)\right). 
\end{eqnarray}
In the real gravitational system, this distribution carries the variation caused by non-local gravitational effects
and many other effects as well. 
For example, one obvious distinguish between our mean trajectories and real evolution is the alignment
between tidal tensor and shear tensor. Since in general, $\langle \varepsilon_{ij} | \vPsi; \tau \rangle$
could be expressed as some function of $\sigma_{ij}$ together with other scalars, their eigenvectors would 
always align with each other.  However, this is not necessarily true in reality.

In the context of halo collapse model, recent studies have already demonstrated the importance of 
incorporating the scattering around those mean trajectories. 
As one of the key ingredients of the Press-Schechter formula \citep{PS74}
or more generally the excursion set theory \citep{BCEK91,Sh98}, the criterion of halo formation is 
manifested as a barrier at which an ensemble of random walks might across. 
In the simplest spherical collapse model, where only the density contrast $\drho$ is considered,  
this criteria corresponds to $\drho = 1.68$.
For homogenous ellipsoidal collapse model, the criterion would depend on the shape as well.
Moreover, it was shown that a more accurate and self-consistent excursion set model would necessarily 
require the barrier to be stochastic \citep{MR10b,CA11a,CA11b,ARSC13}. 
At the single parameter level, i.e. the barrier only depends on $\drho$, the fuzziness of the 
barrier could be seen as the consequence of projecting from higher dimensional dynamical
space of the halo collapse ($\drho$ and the shape) to $\drho$ only. 
Therefore, this is also a self-consistent requirement for HEC. 
But as demonstrated in this paper, it could also arise from other aspects, 
including either spatial variation of the environment or other stochasticities.

Consequently, simply neglecting those randomness and inversely mapping any deterministic 
trajectories will not be able to correctly predict the initial parameter region that eventually leads to 
the collapse. 
From this point of view, our method in this paper only provides the solution at the first level,  i.e.
finding statistically meaningful way to select those mean trajectories. 
But a more accurate description of the problem would require the full knowledge of the 
conditional probability distribution of $\vPsi(\tau_i)$ given the collapsed parameter space 
$\vPsi({\tau})$ at time $\tau$
\begin{eqnarray}
\mP (\vPsi_{\tau_i} | \vPsi_{\tau}) = 
\frac{\mP (\vPsi_{\tau_i}) }  {\mP(\vPsi_{\tau}) } \mP(\vPsi_{\tau} | \vPsi_{\tau_i}) . 
\end{eqnarray}
We will defer the investigation of this probability in the subsequent studies.

\subsection{Conclusion}
In this paper, we presented a statistical method for decoupling the intrinsically non-local 
Lagrangian evolution of a fluid element from a self-gravitational random field in 
Newtonian cosmology. 
Since the gravitational potential is constrained by the Poisson equation, the tidal tensor 
in Newtonian cosmology is highly non-local and determined by all the matter in 
the Universe. 
Therefore, dynamical variables like $\drho$ and $A_{ij}$ of a single fluid element 
could not be uniquely determined by their initial values and would vary spatially.  
Instead of searching for some local approximations \citep{BJ94,HB96}, we ask an alternative 
question, that is, what is the mean fluid evolution with given initial density and shape 
of the element. 
Mathematically, this leads us to the characteristic curves of the transport equation of 
the density-weighted probability density function. 
Physically, if one evolves these local but fully non-linear curves with the same set of initial 
conditions as the real system, it is guaranteed that the one-point $\rho$PDF would 
always be identical.

Besides the tidal tensor, our formalism makes no simplification to the non-linearity of 
the system, therefore it will be very useful for understanding the non-linear evolution of 
dark matter halos as well as other cosmic web morphologies. 
For dark matter halos, or over-dense region in general, the incorporation of internal velocity 
dispersion of the fluid element in the dynamical equation might be helpful for avoiding the 
singularity at the shell-crossing. 
For understanding the evolution of all types of cosmic web morphologies, 
since our method preserves the one-point $\rho$PDF, a direct sampling of all
effective trajectories would provide an accurate estimate of, e.g. the fraction of each
morphological type, for different definitions of the cosmic web.
Moreover, it is also interesting to notice that for primordial non-Gaussian initial condition,
our `localized mean' trajectory obtained here will not be the same as Zel'dovich 
approximation path. However, that does not mean the usual practice for generating non-Gaussian
initial condition is wrong since at those early stage, any second order corrections will be tiny. 

\acknowledgments
The authors sincerely appreciate Michael Wilczek and Charles Meneveau for introducing 
the PDF based method  and its application in turbulence. XW would also like to thank for productive
discussion with Mark Neyrinck, Ue-Li Pen, J. R. Bond, Niayesh Afshordi. 

\appendix
\newcommand{\appsection}[1]{\let\oldthesection\thesection
\renewcommand{\thesection}{\oldthesection}
\section{#1}\let\thesection\oldthesection}

\appsection{Evolution Equation of Density Weighted PDF}
In this Appendix, we will demonstrate the evolution equation of density weighted probability 
distribution function $\mD (\vPsi;\vx,\tau) = (1+\Drho) \mP_E(\vPsi;\vx,\tau)$ at time $\tau$ and 
randomly selected Eulerian position $\vx$. 
Note that we have explicitly expressed the position $\vx$ to
incorporate the convection in the following derivation. 
To proceed, we first consider the extended dynamical system including 
the Eulerian equation of velocity field 
\begin{eqnarray}
\label{eqn:euler_eq}
\frac{d}{d\tau} u_i  = - \mH(\tau) u_i  - \Phi_i . 
\end{eqnarray}
Denoting the extended dynamical variable $\vpsi^{t} = \{\drho, u_i, A_{ij}  \} =  
\{\drho, \theta, u_i, \sigma_{ij}  \} $,  and combining equation (\ref{eqn:euler_eq}) with equation 
(\ref{eqn:dyn_eqn_full}), we will express the full dynamical system as
\begin{eqnarray}
\label{eqn:Lag_eq_stoch}
\frac{d}{d\tau} \vpsi^{t}(\tau)  = \vchi^{t}[\vpsi^{t}, \varepsilon_{ij}; \tau] , 
\end{eqnarray}
where $d/d\tau$ is Lagrangian total derivative. 

Following \cite{Pop85}, we would like to consider the ensemble average of the quantity 
\begin{eqnarray}
\left \langle (1+\drho)  \frac{d}{d\tau} Q(\vpsi^{t}) \right\rangle_E 
\end{eqnarray}
where $Q(\vpsi^{t} )$ is arbitrary function of $\vpsi^{t}$. Here the ensemble average $\langle \cdot \rangle_E$ is taken 
in the Eulerian space at $\vx$ with probability density function $\mP_E(\vPsi^{t}; \vx,\tau )$. 
We then expand the total derivative $d/d\tau$ explicitly
\begin{eqnarray}
\left \langle (1+\drho)  \frac{d}{d\tau} Q \right \rangle_E &=& 
 \left \langle (1+\drho) \left [ \frac{\partial}{\partial \tau} + u_i 
\frac{\partial}{\partial x_i} \right ] Q  \right \rangle_E  \nonumber \\
&=&  \frac{\partial}{\partial \tau} \left \langle  (1+\drho) Q  
\right \rangle_E +  \frac{\partial}{\partial x_i}  \left \langle  (1+\drho) u_i Q \right \rangle_E
 - \left \langle Q \left[ \frac{\partial}{\partial \tau}\drho + 
\frac{\partial}{\partial x_i} [(1+\drho)u_i] \right] \right \rangle_E . 
\end{eqnarray}
In the second equality, we have rearranged all terms so that the last one vanishes due to the continuity 
equation. Since the sample space variables commute with $\partial/\partial \tau$  and $\partial/\partial x_i$, we 
further have 
\begin{eqnarray}
\label{eqn:ePDF_eq1}
\left \langle (1+\drho)  \frac{d}{d\tau} Q \right \rangle_E &=& 
\frac{\partial}{\partial \tau} \int d\vPsi^{t}   \left [  Q(\vPsi^{t}) (1+\Drho)
\mP_E(\vPsi^{t}; \vx, \tau) \right]  +  \frac{\partial}{\partial x_i}  \int d\vPsi^{t} \left [ 
Q(\vPsi^{t}) U_i (1+\Drho) \mP_E(\vPsi^{t}; \vx, \tau)  \right] \nonumber \\
&=&\int d\vPsi^{t} ~  Q(\vPsi^{t}) \left[ \frac{\partial}{\partial \tau} + U_i
\frac{\partial}{\partial x_i} \right ] \mD(\vPsi^{t}; \vx, \tau) , 
\end{eqnarray}
where the dynamical variable $\vpsi^t$ in the sample space is denoted as
$\vPsi^{t} =\{ \Drho, U_i, \mA_{ij} \} $.

On the other hand, one could also derive 
\begin{eqnarray}
\left \langle (1+\drho) \frac{d}{d\tau}Q \right \rangle_E = 
 \left \langle (1+\drho) \frac{d \psi^{t}_{\alpha}}{d\tau} 
\left(\frac{\partial}{\partial \psi^{t}_{\alpha}} Q(\vpsi^{t}) \right) 
\right \rangle_E. 
\end{eqnarray}
Substituting the dynamical equation (\ref{eqn:Lag_eq_stoch}), one has
\begin{eqnarray}
\left \langle (1+\drho) \frac{d}{d\tau}Q \right \rangle_E 
&=& \int d\vPsi^{t}   \left \langle (1+\drho) \chi^{t}_{\alpha} 
\frac{\partial Q}{\partial \psi^{t}_{\alpha}}
 \middle | \vPsi^{t} ;\vx, \tau \right \rangle_E   \mP_E(\vPsi^{t}; \vx, \tau) \nonumber \\
&=& \int d\vPsi^{t} \frac{\partial Q}{\partial\Psi^{t}_{\alpha} } \langle \chi_{\alpha}^{t}
|\vPsi^{t}; \vx, \tau  \rangle_E \mD(\vPsi^{t}; \vx, \tau) .
\end{eqnarray}
Performing Integration by part, one would obtain
\begin{eqnarray}
\label{eqn:ePDF_eq2}
I -  \int d\vPsi^{t}~  Q(\vPsi^{t})\left [ \frac{\partial}{\partial \Psi^{t}_{\alpha}} 
\left( \left \langle \chi_{\alpha}^{t}
\middle | \vPsi^{t} ;\vx,\tau \right \rangle_E   \mD(\vPsi^{t} ;\vx,\tau )  \right) \right]  , 
\end{eqnarray}
where $I$ is the surface integral
\begin{eqnarray}
I = \int d\vPsi^{t}  \frac{\partial}{\partial \Psi^{t}_{\alpha}} \left [ 
Q(\vPsi^{t}) \langle \chi_{\alpha}^{t} | \vPsi^{t} ;\vx,\tau \rangle_E \mD(\vPsi^{t};\vx,\tau )\right], 
\end{eqnarray}
and would vanish for most of functions $Q$ \citep{Pop85}. 
Equating equation (\ref{eqn:ePDF_eq1}) with equation (\ref{eqn:ePDF_eq2}), and
since $Q$ is arbitrary, we obtain the evolution equation of $\mD(\vPsi^{t} ; \vx, \tau )$
\begin{eqnarray}
\frac{\partial}{\partial \tau}\mD(\vPsi^{t};\vx, \tau) +\frac{\partial}{\partial x_i}  U_i \mD(\vPsi^{t}; \vx,\tau)
 + \frac{\partial}{\partial \Psi^{t}_{\alpha}} \langle \chi^{t}_{\alpha} |\vPsi^{t} ;\vx, \tau \rangle_E 
 \mD(\vPsi^{t}; \vx, \tau) =0 . 
\end{eqnarray} 
For statistical homogeneous and isotropic field, $\mP_E$ and $\mD$ do not explicitly 
depend on position $\vx$. Therefore, the term $\partial/\partial x_i (U_i \mD)$ would vanish. 
Furthermore, assuming $\mD$ is bounded in velocity space $U_i$, one could 
further integrate out such contribution so that the equation would only depend on $\vPsi$
instead of $\vPsi^t$.   Eventually, this leads to the equation
\begin{eqnarray}
\frac{\partial}{\partial \tau}\mD(\vPsi; \tau) + \frac{\partial}{\partial \Psi_{\alpha}} 
\langle \chi_{\alpha} |\vPsi ; \tau \rangle_E \mD(\vPsi; \tau)  = 0, 
\end{eqnarray}
which is the same as the evolution equation of Lagrangian PDF $\mP_L(\vPsi; \tau)$.

\appsection{Conditional Average}
\label{eqn:cond_avg}
The major task of the statistical closure method is to estimate the conditional average of the tidal tensor. 
In general, we are interested in the conditional average
of two random vectors $\vx$ and $\vy$, with $\vX$ and $\vY$  corresponding to their sample space 
variable respectively.  Denoting their joint probability function as $\mP(\vX, \vY)$, 
then by definition, the conditional average could be expressed as
\begin{eqnarray}
\label{eqn:cond_avg_def}
\langle \vx | \vY \rangle \mP(\vY) =  \int d \vX ~ \vX ~ \mP(\vX,\vY). 
\end{eqnarray}
In the cosmological context, such quantity have been extensively studied both for Gaussian and weakly non-Gaussian distribution. 
For example, the Gaussian expression of  $ \langle \vx | \vY \rangle$ is well known in cosmology 
since \cite{BBKS86}. 
And for weakly non-Gaussian distributed random variables, the formula here
is very similar to the one adopted in e.g. estimating the Minkowski functional \citep{Mat1994,Mat03,PGP09} . 
To help the reader who are unfamiliar with the subject, we will provide a detailed 
derivation in this appendix.

\subsection{Gaussian Case}
\label{appsec:gaussian}
The standard procedure starts by utilizing the joint characteristic function, defined as the inverse Fourier 
transformation of the probability density function
\begin{eqnarray}
\mZ(\vlambda_x, \vlambda_y)  = \exp\left [ -\frac{1}{2} \left ( \lambda_x^{\alpha}  
\presuper{(2)}{\xi}^{xx}_{\alpha\beta}  
\lambda_x^{\beta} + 2  \lambda_x^{\alpha}   \presuper{(2)}{\xi}^{xy}_{\alpha\beta} \lambda_y^{\beta} 
+ \lambda_y^{\alpha}  \presuper{(2)}{\xi}^{yy}_{\alpha\beta}   \lambda_y^{\beta}  \right)  \right], 
\end{eqnarray}
where $\alpha, \beta \cdots $ are vector indices, and $\presuper{(2)}{\xi}^{xx}_{\alpha\beta}, 
\presuper{(2)}{\xi}^{yy}_{\alpha\beta}, \presuper{(2)}{\xi}^{xy}_{\alpha\beta}$ are covariance matrix 
between vector $\vx-\vx$, $\vy-\vy$ and $\vx - \vy$ respectively.  
Then, writing in the vector form, we have
\begin{eqnarray}
\langle \vx | \vY \rangle \mP(\vY) 
& =& (2\pi)^{-N}  \int d\vX \int d \vlambda_x d\vlambda_y ~ \vX ~
 \exp[ -i  ( \vlambda_x \cdot \vX + \vlambda_y \cdot \vY) ] \mZ(\vlambda_x, \vlambda_y) . 
\end{eqnarray}
After substituting $\vX \exp[  -i  ( \vlambda_x \cdot \vX + \vlambda_y \cdot \vY)]$
with $i \partial /\partial \vlambda_x \exp[  -i  ( \vlambda_x \cdot \vX + \vlambda_y \cdot \vY) $ and 
then performing the integration by part, the $\alpha$ component of the conditional average becomes
\begin{eqnarray}
\langle x_{\alpha} | \vY \rangle \mP(\vY)  &=& 
 (2\pi)^{-N}  \int d \vX \int d \vlambda_x d\vlambda_y ~ \left [ -i \frac{\partial}{\partial \lambda_x^{\alpha} }
 \mZ(\vlambda_x, \vlambda_y)  \right]   \exp[  -i  ( \vlambda_x \cdot \vX + \vlambda_y \cdot \vY)]
\nonumber \\
  &=&   (2\pi)^{-N}  \int d \vX \int d \vlambda_x d\vlambda_y ~  \left [ - \left (
 \presuper{(2)}{\xi}^{xx}_{\alpha\beta} \frac{\partial}{ \partial X_{\beta}} +   
 \presuper{(2)}{\xi}^{xy}_{\alpha\beta} \frac{\partial}{ \partial Y_{\beta}}  \right) 
   \exp[  -i  ( \vlambda_x \cdot \vX + \vlambda_y \cdot \vY)]     \right]  \mZ( \vlambda_x, \vlambda_y ).  \nonumber \\
\end{eqnarray}
Due to the integration over $\vX$, only the term proportional to $\partial / \partial Y_{\beta}$ would 
survive, 
\begin{eqnarray}
\label{eqn:cond_avg_gauss_derv}
\langle x_{\alpha} | \vY \rangle \mP(\vY)  &=& -  \presuper{(2)}{\xi}^{xy}_{\alpha\beta} \left ( 
\frac{\partial}{\partial  Y_{\beta}}  \mP(\vY) \right).  
\end{eqnarray}
Therefore, the conditional average simply reads
\begin{eqnarray}
\label{eqn:cond_avg_gauss}
\langle x_{\alpha} | \vY \rangle =  \presuper{(2)}{\xi}^{xy}_{\alpha\beta} \left (\presuper{(2)}{\xi}^{yy} 
\right) ^{-1} _{\beta\gamma}
Y_{\gamma}, 
\end{eqnarray}
where the inverse of covariance matrix $ \left ( \presuper{(2)}{\xi}^{yy} \right ) ^{-1} _{\beta\gamma}$ arises from
the derivative of the Gaussian probability density function $\mP(\vY)$.

\subsection{Weakly Non-Gaussian Case}
\label{appsec:nongaussian}

For weakly non-Gaussian field, we could apply the cumulant expansion theorem and expand an 
arbitrary distribution function $\mP$  in terms of $n-$th order of cumulants \citep{Cra1946,Ken1958,Mat1994,Mat03,PGP09}.
\begin{eqnarray}
\mP(\vY) = \exp\left [  \sum_{n\ge 3} \frac{(-1)^n}{n!} \presuper{(n)}{\xi}^{y}_{\alpha_1\cdots \alpha_n} 
\frac{\partial ^n }{ \partial Y_{\alpha_1} \cdots \partial Y_{\alpha_n}}  \right] \mP_G(\vY), 
\end{eqnarray}
where $\mP_G(\vY)$ is Gaussian distribution of $\vY$, $\presuper{(n)}{\xi}^{y}_{\alpha_1\cdots \alpha_n} $
 is $n-$point cumulant function of variable $\vY$. 
Starting from equation (\ref{eqn:cond_avg_def}), we need to expand both $\mP(\vY)$  
as well as the joint PDF $\mP(\vX, \vY)$. Keep to the next leading order, and define new
vector $\vGamma=\{\vX, \vY\}$. 
\begin{eqnarray}
\label{eqn:condavg_bispec_def}
\langle x_{\alpha} | \vY \rangle& =&  \left [  1+ \frac{1}{3!} \mP^{-1}_G (\vY) 
  \presuper{(3)}{\xi}^{yyy}_{\beta\gamma\delta} \left ( \frac{\partial^3}{\partial Y_{\beta} \partial Y_{\gamma} 
  \partial Y_{\delta} } \mP_G(\vY)   \right)   \right]  \langle x_{\alpha} | \vY \rangle_G \nonumber \\
&&  -  \frac{1}{3!} \mP^{-1}_G(\vY) \presuper{(3)}{\xi}^{\Gamma\Gamma\Gamma}_{\beta\gamma\delta}
  \int d\vX ~ X_{\alpha}  ~ \left ( \frac{\partial^3}{\partial \Gamma_{\beta}  \partial \Gamma_{\gamma}
  \partial \Gamma_{\delta} } \mP_G(\vGamma) \right ), 
\end{eqnarray}
where $\langle x_{\alpha}| \vY \rangle_G$ is the conditional average regarding the Gaussian distribution
of $\mP_G$. 
For the second term, the third order derivative of $\mP(\vGamma)$ with respect to $\vGamma$ could 
either be $\vX$ or $\vY$.  
However, due to the integration over $\vX$, the derivative with respect to $\vX$ could be no more than 
first order. So the second term could be reduced to
\begin{eqnarray}
&& -  \frac{1}{3!} \mP^{-1}_G(\vY) \presuper{(3)}{\xi}^{\Gamma\Gamma\Gamma}_{\beta\gamma\delta}
  \int d\vX ~ X_{\alpha}  ~ \left ( \frac{\partial^3}{\partial \Gamma_{\beta}  \partial \Gamma_{\gamma}
  \partial \Gamma_{\delta} } \mP_G(\vGamma) \right ) \nonumber\\
&=& - \frac{1}{3!} \mP^{-1}_G(\vY)   \int d\vX ~ X_{\alpha} 
\left ( \presuper{(3)}{\xi}^{yyy}_{\beta\gamma\delta}   \frac{\partial^3}{\partial Y_{\beta}  \partial Y_{\gamma}  
\partial Y_{\delta} }   + 3  \presuper{(3)}{\xi}^{xyy}_{\beta\gamma\delta}  
 \frac{\partial^3}{\partial X_{\beta}  \partial Y_{\gamma}  
\partial Y_{\delta} }   \right )   \mP_G(\vGamma) .
\end{eqnarray}
After substituting the definition equation (\ref{eqn:cond_avg_def}) and performing the integration by 
part that eliminates $X_{\alpha}$, this term would become
\begin{eqnarray}
   -  \frac{1}{3!} \mP^{-1}_G(\vY)  \left [    \presuper{(3)}{\xi}^{yyy}_{\beta\gamma\delta}  
  \frac{\partial^3}{\partial Y_{\beta}  \partial Y_{\gamma}  \partial Y_{\delta} }  \left( \langle x_{\alpha}
  | \vY \rangle_G \mP_G(\vY) \right) - 3  \presuper{(3)}{\xi}^{xyy}_{\alpha\beta\gamma}  
 \frac{\partial^2}{\partial Y_{\beta} \partial Y_{\gamma} }  \mP_G(\vGamma)  \right ]. 
\end{eqnarray}
One then notices the cancellation with the first term in equation (\ref{eqn:condavg_bispec_def}), 
and becuase the Gaussian conditional average $\langle x_{\alpha} | \vY \rangle_G$ is
 linearly proportional to $\vY$, one further reduces to
\begin{eqnarray}
\langle x_{\alpha} | \vY \rangle& =&   \langle x_{\alpha} | \vY \rangle_G
+ \frac{1}{2} \mP^{-1}_G(\vY) \left ( \frac{\partial^2}{\partial Y_{\beta} \partial Y_{\gamma}} 
\mP_G(\vY) \right) \left[    \presuper{(3)}{\xi}^{xyy}_{\alpha\beta\gamma} 
-  \presuper{(3)}{\xi}^{yyy}_{\beta\gamma\delta}  \left( \frac{\partial }{ \partial Y_{\delta}}   
  \langle x_{\alpha} | \vY\rangle_G \right)  \right]. 
\end{eqnarray}
Since the second derivative of Gaussian PDF could be expressed as two-point covariance 
matrix
\begin{eqnarray}
\mP^{-1}_G(\vY)  \left ( \frac{\partial^2}{\partial Y_{\beta} \partial Y_{\gamma}} \mP_G(\vY) \right)
= - \left( \presuper{(2)}{\xi}^{yy} \right)^{-1}_{\beta\gamma} 
+  \left( \presuper{(2)}{\xi}^{yy} \right)^{-1}_{\beta\lambda}  
 \left( \presuper{(2)}{\xi}^{yy} \right)^{-1}_{\gamma\tau}  Y_{\lambda} Y_{\tau}, 
\end{eqnarray}
Eventually, we could expressed the conditional average as
\begin{eqnarray}
\label{eqn:cond_avg_skew}
\langle x_{\alpha} | \vY \rangle& =&   \langle x_{\alpha} | \vY \rangle_G
+ \frac{1}{2} \left [   \presuper{(3)}{\xi}^{xyy}_{\alpha\beta\gamma} 
-  \presuper{(3)}{\xi}^{yyy}_{\beta\gamma\delta}  \presuper{(2)}{\xi}^{xy}_{\alpha\kappa}
\left ( \presuper{(2)}{\xi}^{yy}  \right )^{-1}_{\kappa\delta}   \right] 
 \left [  - \left( \presuper{(2)}{\xi}^{yy} \right)^{-1}_{\beta\gamma} 
+  \left( \presuper{(2)}{\xi}^{yy} \right)^{-1}_{\beta\lambda}  
 \left( \presuper{(2)}{\xi}^{yy} \right)^{-1}_{\gamma\tau}  Y_{\lambda} Y_{\tau}  \right ]. 
\nonumber \\
\end{eqnarray}

\appsection{Cumulant Matrices}
\label{appsec:cumulant_matrix}

In this section, we express all elements of both two- and three-point correlation function that are 
needed for calculating the conditional average. 
In the following of the section, we will not explicitly denote the order number, since it could be
inferred directly from the number of indices.

\subsection{Covariance Matrices $\xi^{\psi\psi}$, $\xi^{\varepsilon\psi}$}
\label{appsec:covmat}
In this subsection, we will list all components of the covariance matrix between $\vpsi$ and 
$\varepsilon_{ij}$. 
Since the inverse of the matrix $\xi^{\psi\psi}$ will be singular if one decompose 
$\vpsi=\{\drho, \theta, \sigma_{ij} \}$, we will instead calculate the covariance matrix between 
$\vpsi=\{\drho, A_{ij} \}$
\begin{eqnarray}
\xi^{\delta\delta} = \sigma^2_{\delta\delta} , \qquad
\xi^{\delta A}_{ij} = \frac{\sigma^2_{\delta\theta}}{3} \delta^K_{ij}, \qquad
\xi^{AA}_{ij,mn} = \frac{\sigma^2_{\theta\theta}}{15} \left( \delta^K_{ij}\delta^K_{mn} + 
\delta^K_{im}\delta^K_{jn} + \delta^K_{in}\delta^K_{jn} \right), 
\end{eqnarray}
where we have defined the variances
\begin{eqnarray}
\sigma^2_{ab} = \int \frac{d^3k}{(2\pi)^3} P_{ab}(k)
\end{eqnarray}
$a, b = \{ \delta, \theta\}$. 
We are also interested in the inverse of $\xi^{\psi\psi}$, 
\begin{eqnarray}
\label{eqn:xipsipsi_inv}
(\xi^{-1}_{\psi\psi})^{\delta\delta} &=& D_1 = -\frac{\sigma^2_{\theta\theta}}{ \sigma^4_{\delta\theta} - 
                                         \sigma^2_{\delta\delta}\sigma^2_{\theta\theta}} \nonumber \\
(\xi^{-1}_{\psi\psi})^{\delta A}_{ij} &=& D_2 \delta^K_{ij}  = \frac{\sigma^2_{\delta\theta}}{ \sigma^4_{\delta\theta} - 
                                         \sigma^2_{\delta\delta}\sigma^2_{\theta\theta}}  \delta^K_{ij} \nonumber \\
(\xi^{-1}_{\psi\psi})^{A A}_{ij,mn} &=& \frac{D_3}{2} \left(\delta^K_{im}\delta^K_{jn} + \delta^K_{in}\delta^K_{jm} \right) 
                       + D_4 \delta^K_{ij}\delta^K_{mn}   \nonumber\\
   &=& \frac{15}{4\sigma^2_{\theta\theta} }  \left(\delta^K_{im}\delta^K_{jn} + \delta^K_{in}\delta^K_{jm} \right) 
 -\frac{1}{2\sigma^2_{\theta\theta}}  \left( \frac{ 5 \sigma^4_{\delta\theta} - 3 \sigma^2_{\delta\delta} 
 \sigma^2_{\theta\theta}}  { \sigma^4_{\delta\theta} - \sigma^2_{\delta\delta}\sigma^2_{\theta\theta} } \right)
  \delta^K_{ij}\delta^K_{mn} .
\end{eqnarray}
And we will also define 
\begin{eqnarray}
D_5 = D_4 + \frac{1}{3}D_3 =   -\frac{\sigma^2_{\delta\delta}}{ \sigma^4_{\delta\theta} - 
        \sigma^2_{\delta\delta}\sigma^2_{\theta\theta}}.
\end{eqnarray}

On the other hand, since for $\vpsi_{\alpha} = \drho$, the covariance vanishes 
$\xi^{\varepsilon \delta}_{ij} = 0$, the only contribution of the covariance matrix 
$\xi^{\varepsilon \psi}$ is then
\begin{eqnarray}
\xi^{\varepsilon A}_{ij,mn} = \frac{4\pi  G_N \bar{\rho} a^2 }{45}\sigma^2_{\delta\theta} 
\left(  3 \delta^K_{im}\delta^K_{jn} + 3 \delta^K_{in}\delta^K_{jm} - 2 \delta^K_{ij}\delta^K_{mn} \right)
\end{eqnarray}

\subsection{Coskewness Matrices $\xi^{\sigma\psi\psi}$, $\xi^{\varepsilon\psi\psi}$}
\label{appsec:cskewmat}

For non-Gaussian closure, we also need the third order cumulants 
$\xi^{\sigma\psi\psi}$ and $\xi^{\varepsilon\psi\psi}$, where, similar to covariance matrix, we will 
parametrize $\vpsi=\{ \drho, A_{ij}\}$. 
First of all, it is easy to show the following components vanish
\begin{eqnarray}
\xi^{\varepsilon\delta\delta}_{ij} = \xi^{\sigma \delta\delta}_{ij} = 0.
\end{eqnarray}

For  component between $\varepsilon_{ij}, \drho$ and $A_{mn}$, 
\begin{eqnarray}
\xi^{\varepsilon \delta A}_{ij, mn} &= &\xi^{\varepsilon \delta A} \left (
 3 \delta^K_{im}\delta^K_{jn} + 3\delta^K_{in} \delta^K_{jm} - 2 \delta^K_{ij} \delta^K_{mn}  \right ) 
 , \qquad where \nonumber\\
  \xi^{\varepsilon \delta A} &= & (4\pi G_N \bar{\rho} a^2) \int \frac{d\vk_{123}}{(2\pi)^6} 
  \left ( \frac{3\mu_{13}^2 - 1}{90} \right) \delta_D(\vk_{123}) B^{\delta\delta \theta} (\vk_1, \vk_2, \vk_3). 
\end{eqnarray}
Where $B^{\delta\delta\theta}(\vk_1, \vk_2, \vk_3)$ is the bispectrum between $\delta, \delta$ and $\theta$, 
$\mu_{13} $ is the cosine of the angle between $\vk_1$ and $\vk_3$. 
And we have denoted the $\delta_D(\vk_{123}) =  \delta_D(\vk_1+\vk_2 + \vk_3)$ explicitly in the integration
and written the volume element $d\vk_{123} = d\vk_1 d\vk_2 d\vk_3 = k_1^2 k_2^2 k_3^2 
d\vOmega_{k1} d\vOmega_{k2} d\vOmega_{k3} $. 
Similarly, 
\begin{eqnarray}
\xi^{\sigma \delta A}_{ij, mn} &= &\xi^{\sigma \delta A} \left (
 3 \delta^K_{im}\delta^K_{jn} + 3\delta^K_{in} \delta^K_{jm} - 2 \delta^K_{ij} \delta^K_{mn}  \right ) 
 , \qquad and \nonumber\\
  \xi^{\sigma \delta A} &= &  \int \frac{d\vk_{123}}{(2\pi)^6} 
  \left ( \frac{3\mu_{13}^2 - 1}{90} \right) \delta_D(\vk_{123}) B^{\theta\delta \theta} (\vk_1, \vk_2, \vk_3). 
\end{eqnarray}

Again, the matrix components of $(\varepsilon_{ij},A_{mn}, A_{kl}$) and $(\sigma_{ij}, A_{mn}, A_{kl}$)
are very similar. 
\begin{eqnarray}
\xi^{\varepsilon/\sigma A A}_{ij,mn,kl} &=& \xi^{\varepsilon/\sigma A A}_1 ~ \delta^K_{ij} \delta^K_{mn}\delta^K_{kl}
+ \xi^{\varepsilon/\sigma A A}_2 \left (  \delta^K_{ij} \delta^K_{mk}\delta^K_{nl} +
\delta^K_{ij}\delta^K_{ml}\delta^K_{nk} \right) + \xi^{\varepsilon/\sigma A A}_3 \bigl( 
\delta^K_{mn}\delta^K_{ik}\delta^K_{jl} + \delta^K_{mn}\delta^K_{il}\delta^K_{jk}  \nonumber \\
&& + \delta^K_{kl}\delta^K_{im}\delta^K_{jn} + \delta^K_{kl}\delta^K_{in}\delta^K_{jm}  \bigr)
+ \xi^{\varepsilon/\sigma A A}_4 \bigl( \delta^K_{im} \delta^K_{jk}\delta^K_{nl} + 
\delta^K_{im}\delta^K_{jl}\delta^K_{kn} + \delta^K_{in}\delta^K_{jk}\delta^K_{ml} + 
\delta^K_{in}\delta^K_{jl}\delta^K_{km}   \nonumber \\
&& + \delta^K_{ik}\delta^K_{jm}\delta^K_{nl} + \delta^K_{ik}\delta^K_{jn}\delta^K_{ml}
+ \delta^K_{il}\delta^K_{jm}\delta^K_{nk} + \delta^K_{il}\delta^K_{jn}\delta^K_{mk}  \bigr). 
\end{eqnarray}
where the coefficient could be written separately
\begin{eqnarray}
\xi^{\varepsilon A A}_i &=& (4\pi G_N \bar{\rho} a^2) \int \frac{d\vk_{123}}{(2\pi)^6} 
 ~ \mK_A^{(i)}  ~\delta_D(\vk_{123}) B^{\delta \theta \theta} (\vk_1, \vk_2, \vk_3) \nonumber \\
\xi^{\sigma A A}_i &=& \int \frac{d\vk_{123}}{(2\pi)^6} 
 ~ \mK_A^{(i)}  ~\delta_D(\vk_{123}) B^{\theta \theta \theta} (\vk_1, \vk_2, \vk_3). 
\end{eqnarray}
Here $1\le i \le 4$, and the angular kernel $\mK_A^{(i)}$ equals
\begin{eqnarray}
\mK_A^{(1)}= \frac{2}{315} \left( 5 - 15 \mu^2_{13} + 12 \mu_{12}\mu_{13} \mu_{23} - 
                           4 \mu^2_{23} \right) , \qquad
 \mK_A^{(2)}= \frac{2}{315} \left( -2 + 6 \mu^2_{13} -9 \mu_{12}\mu_{13} \mu_{23} + 
                           3 \mu^2_{23} \right) \nonumber \\
\mK_A^{(3)}= \frac{1}{210} \left( -5 + 15 \mu^2_{13} - 12 \mu_{12}\mu_{13} \mu_{23} +
                           4 \mu^2_{23} \right), \qquad
\mK_A^{(4)}= \frac{1}{210} \left( 2 - 6 \mu^2_{13} + 9 \mu_{12}\mu_{13} \mu_{23} -
                           3 \mu^2_{23} \right) . \nonumber\\
\end{eqnarray}
As it turned out, some combinations of $\xi^{\varepsilon/\sigma A A}_i$ will also be very useful, which
we define as
\begin{eqnarray}
\tilde{\xi}^{\varepsilon/\sigma A A}_1 &=& \delta^K_{mn}\delta^K_{ik}\delta^K_{jl} 
\xi^{\varepsilon/\sigma A A}_{ij,mn,kl} = 9 \xi^{\varepsilon/\sigma A A}_1 + 6
\xi^{\varepsilon/\sigma A A}_2 + 42 \xi^{\varepsilon/\sigma A A}_3 + 48 \xi^{\varepsilon/\sigma A A}_4, 
\nonumber\\
\tilde{\xi}^{\varepsilon/\sigma A A}_2 &=& \delta^K_{im}\delta^K_{jk}\delta^K_{nl} 
\xi^{\varepsilon/\sigma A A}_{ij,mn,kl} = 3 \xi^{\varepsilon/\sigma A A}_1 + 12
\xi^{\varepsilon/\sigma A A}_2 + 24 \xi^{\varepsilon/\sigma A A}_3 + 66 \xi^{\varepsilon/\sigma A A}_4. 
\end{eqnarray}
And their angular kernels are
\begin{eqnarray}
\tilde{\mK}_A^{(1)} & = & \mu^2_{13} - \frac{1}{3} \nonumber\\ 
\tilde{\mK}_A^{(2)} & = & \mu_{12}\mu_{13} \mu_{23} - \frac{1}{3} \mu^2_{23} . 
\end{eqnarray}

\appsection{The Gaussian Closure of Tidal Tensor in the Integration Form}
\label{appsec:gaussclos_intg}

\subsection{Derivation}
\label{appsec:int_derv}
In order to have a better understanding the non-locality of tidal tensor, in this section, 
we will perform the same statistical closure of the $\varepsilon_{ij}$ in the integral form.  
Given the integral solution of the peculiar gravitational potential \citep{Bert95}, 
\begin{eqnarray}
\Phi (\vx) =  - G_N \bar{\rho} a^2 \int d^3 \vx^{\pr} ~ \frac{\drho(\vx^{\pr})}{r}  
\end{eqnarray}
where $\vrr = \vx^{\pr}-\vx$, the peculiar tidal tensor could be expressed as 
\citep{C66,OK95,WM14}
\begin{eqnarray}
\varepsilon_{ij} (\vx) = G_N \bar{\rho} a^2 \int_{P.V.} d^3x^{\pr} \left [ 
\frac{\delta^K_{ij}}{r^3} - 3 \frac{r_i r_j }{r^5} \right] \drho(\vx^{\pr}) .
\end{eqnarray}
Here the integration is taken in the sense of principal value. 
Denoting the kernel within the bracket as $K_{ij}(\vx^{\pr},\vx)$, we then would like to 
evaluate the conditional average
\begin{eqnarray}
\langle \varepsilon_{ij}  | \vPsi; \vx, \tau \rangle  = 
G_N\bar{\rho} a^2 \int_{P.V.} d^3x^{\pr} ~ K_{ij}(\vx^{\pr},\vx)  ~
\langle \drho(\vx^{\pr}) | \vPsi (\vx) \rangle . 
\end{eqnarray}
Therefore, instead of local distribution of $\mP(\vPsi, \vE_{ij})$, 
we have to assume the Gaussian distributed two-point joint probability distribution function 
$\mP^{(2)}(\vPsi_1, \vPsi_2) = \mP^{(2)}(\vPsi(\vx_1),  \vPsi(\vx_2))$. 
By definition, the conditional average of density perturbation $\drho$ at position $\vx_2$, 
given $\vPsi$ at $\vx_1$, could be expressed as
\begin{eqnarray}
\langle \delta_{\rho,2} | \vPsi_1 \rangle \mP^{(1)}(\vPsi_1) &=& \int d\vPsi_2
 ~ \Delta_{\rho,2}~ \mP^{(2)}(\vPsi_1,\vPsi_2) , 
\end{eqnarray}
where $\mP^{(1)}(\vPsi)$ is the one-point PDF. 
Similarly, from equation (\ref{eqn:cond_avg_gauss}), we obtain
\begin{eqnarray}
\langle \delta_{\rho,2} | \vPsi_1 \rangle = \xi^{\delta\psi}_{\alpha}(\vrr) 
\left [ \xi^{\psi\psi}(0) \right]^{-1}_{\alpha\beta} \Psi_{1,\beta}, 
\end{eqnarray}
where $\xi^{\delta\psi}_{\alpha}(\vrr) $ is the correlation function between $\drho$ and $\vpsi$
at distance $\vrr$, and $ \left [ \xi^{\psi\psi}(0) \right]^{-1}_{\alpha\beta} $ are the same as components 
shown in equation (\ref{eqn:xipsipsi_inv}). 
After substituting explicitly the correlation function $\xi^{\delta\psi}_{\beta}$ as shown in 
Appendix \ref{appsec:int_sta}, the conditional average could be expressed as 
\begin{eqnarray}
\langle \drho(\vx^{\pr}) | \vPsi(\vx) \rangle &=& \xi_{\delta}(r) \Drho(\vx) + 
\xi_{\theta}(r) \Theta(\vx)   + \xi_{A}(r) \mA_{ij}(\vx) \hr_i \hr_j. 
\end{eqnarray}
Various scale independent functions $\xi(r)$ depend on correlation functions
\begin{eqnarray}
\xi_{\delta}(r) &=& D_1 \xi^{\delta\delta}(r) + 2 D_2 \xi^{\delta A}_{\perp} (r)
+ D_2 \xi^{\delta A}_{\parallel} (r) \nonumber \\
\xi_{\theta}(r) &=& D_2 \xi^{\delta\delta}(r) + (2 D_4 + D_3) \xi^{\delta A}_{\perp}(r) + 
  D_4 \xi^{\delta A}_{\parallel}(r)   \nonumber \\
\xi_A(r) &=& D_3 \left ( \xi^{\delta A}_{\parallel}(r) - \xi^{\delta A}_{\perp}(r) \right), 
\end{eqnarray}
where $D_1 \sim D_4$ are the same quantities defined in equation (\ref{eqn:xipsipsi_inv}), and
$\xi^{\delta A}_{\parallel} $ and $\xi^{\delta A}_{\perp}$ are parallel and transverse 
component of $\drho - A_{ij}$ correlation function $\xi^{\delta A}_{ij}(\vrr)$.   
As shown in Appendix \ref{appsec:int_sta}, the angular integration of the kernel $K_{ij}$ itself 
vanishes, and the only term would not be zero is the one proportional to $\hr_i \hr_j$ (equation 
\ref{eqn:Krr_integral}).  Therefore, this leads to
\begin{eqnarray}
\langle \varepsilon_{ij}  | \vPsi ; \vx, \tau \rangle &=&  - \frac{8\pi}{ 5}  
G_N \bar{\rho} a^2 D_3
\left (  \int_0^{\infty} dr ~ \frac{\xi_A(r)}{r} \right) \Sigma_{ij}(\vx) . \nonumber \\
\end{eqnarray}
The explicite evaluation of radial integration would give $-\sigma^2_{\delta\theta}/3$. 
Eventually, we obtain 
\begin{eqnarray}
\label{eqn:eij_condition}
\langle \varepsilon_{ij}| \vPsi; \vx, \tau\rangle = \frac{4\pi G_N\bar{\rho} a^2 
\sigma^2_{\delta\theta} }{\sigma^2_{\theta\theta}} ~ \Sigma_{ij} (\vx), 
\end{eqnarray}
so we recover equation (\ref{eqn:eij_gauss_clos}).

\subsection{Two-point Statistics and Angular Integration of Kernel}
\label{appsec:int_sta}
In this subsection, we display all relevant formula needed for estimating the conditional average
in section \ref{appsec:int_derv}. 
Denoting the correlation between $\vpsi = \{ \drho, A_{ij} \}$ as 
$\xi^{\psi\psi}_{\alpha\beta}(\vrr)$, we could write down each components 
\begin{eqnarray}
\xi^{\delta\delta} (r) =  \sigma_{\delta}^2 = \int \frac{dk}{2\pi^2} k^2 ~
P_{\delta\delta}(k) j_0(kr) , 
\end{eqnarray}
where $j_n$ is $n-$th order spherical Bessel function. The correlation between 
$\drho$ and $A_{ij}$ could be decomposed as 
\begin{eqnarray}
\xi^{\delta A}_{ij} (\vrr) &=& \langle \delta(\vx) A_{ij}(\vx+\vrr) \rangle 
= \xi^{\delta A}_{\perp}(r)\delta^K_{ij} +  
[ \xi^{\delta A}_{\perp}(r) - \xi^{\delta A}_{\parallel}(r)]  \hr_i\hr_j , \quad where \nonumber \\
\xi^{\delta A}_{\parallel}(r) &=& \frac{1}{2\pi^2} \int dk k^2 ~ P_{\delta \theta} (k)
\left [ j_0(kr) - 2 \frac{j_1(kr)}{kr} \right] \nonumber\\
\xi^{\delta A}_{\perp}(r) &=& \frac{1}{2\pi^2} \int dk k^2 P_{\delta \theta}(k) ~
~ \frac{j_1(kr)}{kr}. 
\end{eqnarray}
Similarly, for correlation between $A_{ij}$, we have
\begin{eqnarray}
\xi^{AA}_{ij,mn} (\vrr) &=& \langle A_{ij}(\vx) A_{mn}(\vx+\vrr) \rangle \nonumber\\
&=& \frac{1}{8}[ \xi^{AA}_1(r) - 2\xi^{AA}_2(r) + \xi^{AA}_3(r) ] 
( \delta^K_{ij}\delta^K_{mn} + \delta^K_{im}\delta^K_{jn} 
 + \delta^K_{in}\delta^K_{jm}) + \frac{1}{8}[  - \xi^{AA}_1(r) +
 6\xi^{AA}_2(r) -5\xi^{AA}_3(r) ] \nonumber \\
&&\times ( \hr_i\hr_j \delta^K_{mn} + \hr_m\hr_n\delta^K_{ij} + \hr_i\hr_m\delta^K_{jn}
+ \hr_i\hr_n \delta^K_{jm}  + \hr_j\hr_m\delta^K_{in} + \hr_j\hr_n\delta^K_{im}) + 
\frac{1}{8}[ 3\xi^{AA}_1(r) -30 \xi^{AA}_2(r)   \nonumber \\
&& + 35\xi^{AA}_3(r) ] \hr_i\hr_j\hr_m\hr_n . 
\end{eqnarray}
where three independent components 
\begin{eqnarray}
\xi^{AA}_1(r) &=& \int  \frac{dk}{2\pi^2} k^2 P_{\theta\theta}(k)  ~j_0(kr) \nonumber\\
\xi^{AA}_2(r) &=& \int  \frac{dk}{2\pi^2} k^2 P_{\theta\theta}(k)
\left [  j_0(kr) - \frac{2 j_1(kr)}{kr} \right] \nonumber\\
\xi^{AA}_3 (r) &=& \int \frac{dk}{2\pi^2} k^2 P_{\theta\theta}(k) \left[ 
\frac{8 j_2(kr)}{k^2r^2} - \frac{4 j_1(kr)}{kr} + j_0(kr) \right] .  \nonumber\\
\end{eqnarray}

To proceed, we also need to evaluate some integrals related to the kernel
$K_{ij} = \frac{ \delta^K_{ij} - 3 \hr_i \hr_j  }{r^3}  = \frac{ K^{\Omega}_{ij}}{r^3} $. 
First we notice that, the integration of  
\begin{eqnarray}
\int d\Omega ~ \hr_i \hr_j = \frac{4\pi}{3} \delta^K_{ij}. 
\end{eqnarray}
Therefore the integral of $K^{\Omega}_{ij}$ would vanish
\begin{eqnarray}
\int d\Omega~ K^{\Omega}_{ij} = \int d\Omega ~ \left ( \delta^K_{ij} - 3 \hr_i \hr_j \right)   = 0.
\end{eqnarray}
Then we are interested in the integration $\int d\Omega ~ K^{\Omega}_{ij} ~\hr_m \hr_n$. 
From the symmetric consideration, we have 
\begin{eqnarray}
\int d\Omega ~ \hr_i\hr_j\hr_m\hr_n  
 =   q \left [ \delta^K_{ij}\delta^K_{mn} + \delta^K_{im}\delta^K_{jn}
+\delta^K_{in}\delta^K_{jm} \right].  
\end{eqnarray}
Contracting indices with $\delta^K_{ij} \delta^K_{mn}$, we could obtain $q=4\pi/15$. 
Therefore, one has
\begin{eqnarray}
\label{eqn:Krr_integral}
\int d\Omega ~ K^{\Omega}_{ij} ~  \hr_m \hr_n \mA_{mn} =
\int d\Omega \left(\delta^K_{ij}\hr_m \hr_n -3 \hr_i\hr_j\hr_m\hr_n \right)
\mA_{mn}
  = - \frac{8\pi}{5} \left [ \mA_{ij} - \frac{ \delta^K_{ij}}{3} \mA_m^m \right]. 
\end{eqnarray}

\label{lastpage}

\end{document}